\documentclass[aps,physrev,reprint,superscriptaddress,longbibliography,floatfix]{revtex4-2}

\usepackage{amsmath,amssymb,bm,mathtools}
\usepackage{graphicx}
\graphicspath{{figures/}}
\usepackage{microtype}
\usepackage{booktabs}
\usepackage{array}
\usepackage{tabularx}
\usepackage{xcolor}
\usepackage{xurl}
\usepackage{etoolbox}
\usepackage{tikz}
\usepackage{placeins}
\usepackage{hyperref}
\usepackage{bookmark}
\hypersetup{hidelinks}
\usetikzlibrary{
  arrows.meta,
  positioning,
  fit,
  calc,
  shapes.geometric,
  shapes.symbols
}
\newcommand{\qRM}{\mathrm{qRM}}
\newcommand{\St}{\mathrm{St}}
\newcommand{\ket}[1]{\lvert #1\rangle}
\newcommand{\bra}[1]{\langle #1\rvert}
\newcommand{\proj}[1]{\ket{#1}\!\bra{#1}}
\newcommand{\logical}[1]{\overline{#1}}

\newcommand{\GF}{\mathbb F_2}
\makeatletter
\AtBeginDocument{\expandafter\gdef\csname b@apsrev42Control\endcsname{{0}{0}{{}}{{}}}}
\makeatother

\newcommand{\PaperDate}{\today}

\begin{document}

\title{Extremely Low-Cost Magic State Preparation toward Fault-Tolerant Quantum Computing}

\author{Jianshuo Gao}
\affiliation{School of Physics, Peking University, Beijing 100871, China}

\author{Xiao Yuan}
\affiliation{Center on Frontiers of Computing Studies, School of Computer Science, Peking University, Beijing 100871, China}

\author{Yuan Yao}
\affiliation{Center on Frontiers of Computing Studies, School of Computer Science, Peking University, Beijing 100871, China}

\date{\PaperDate}

\begin{abstract}
Fault-tolerant preparation of non-Clifford resource states is a major contributor to the overhead of quantum computation, motivating protocols that achieve high output fidelity with minimal qubit and circuit costs. We introduce a low-cost magic-state preparation protocol in which the choice of stabilizer generators is co-designed with the flag gadgets, allowing the syndrome-extraction circuit itself to filter correlated faults across a non-Clifford layer. The protocol prepares a logical plus state in the 15-qubit quantum Reed–Muller code, applies a transversal \(T\) gate, and gauge-fixes the same register into the seven-qubit Steane code. By reorganizing equivalent \(Z\)-type stabilizer generators into jointly flagged measurement groups, the protocol eliminates all accepted logical-error contributions arising from one or two circuit faults under destructive error detection. Under a uniform circuit-level depolarizing noise model, the postselected infidelity is $210.2p^3+O(p^4)$. 
At \(p=10^{-3}\), exact low-order enumeration combined with stratified sampling bounds the infidelity by \(2.2\times10^{-7}\) at 99.9\% joint confidence, while retaining an acceptance probability of 86.9\%. The complete circuit requires only 19 qubits and 82 CNOT gates. These results demonstrate that stabilizer-generator design can substantially reduce the cost of postselected magic-state preparation, although corrected operation and the fidelity of an unmeasured output block require separate analysis.
\end{abstract}

\maketitle

\section{Introduction}
\label{sec:introduction}

Fault-tolerant quantum computation requires a universal set of logical operations. However, the Eastin–Knill theorem shows that a universal gate set cannot be realized solely through transversal encoded operations~\cite{eastin2009restrictions}. This limitation motivated the development of alternative approaches for implementing non-Clifford operations within fault-tolerant architectures. Among these approaches, the Clifford+\(T\) framework became particularly influential because Clifford operations are efficiently supported by many quantum error-correcting codes, while the \(T\) gate provides a minimal non-Clifford extension that enables universal quantum computation. Moreover, several important quantum codes, such as the 15-qubit quantum Reed–Muller code, allow fault-tolerant implementations of the \(T\) gate through transversal constructions. This combination led to the widespread adoption of the magic-state approach, in which an encoded \(T\) state, \(\ket{T}=T\ket{+}\), is prepared as a consumable resource for universal computation~\cite{bravyi2005universal}. 

The high overhead of preparing high-fidelity magic states remains a major challenge for fault-tolerant quantum computing. Conventional approaches rely primarily on magic-state distillation, where noisy non-Clifford resources are purified using Clifford operations and measurements~\cite{bravyi2005universal}. Subsequent developments have improved the efficiency of distillation through structured code constructions, multilevel protocols, and optimized fault-tolerant factory architectures~\cite{bravyi2012magic,jones2013multilevel,litinski2019magic,gidney2019efficient}. Nevertheless, the resource cost of these approaches remains strongly dependent on the target fidelity and underlying code architecture~\cite{beverland2021cost}. This has motivated alternative encoded preparation strategies, including magic-state cultivation and code-switching approaches, which exploit code structure and logical evolution to reduce resource requirements~\cite{gidney2024magic,daguerre2025code}. 

Efficient error detection is a common ingredient in these preparation schemes. Flag-based methods introduce auxiliary flag qubits to identify correlated faults with low ancilla overhead~\cite{chamberland2019magic,chamberland2020very,yenilen2026sqrtT}, while fault-tolerant postselection exploits syndrome information to suppress accepted logical errors~\cite{bombin2024fault}. These techniques suggest that the structure of error-detection circuits itself can provide additional resources for reducing magic-state preparation overhead.

The $[[15,1,3]]$ quantum Reed--Muller (qRM) code and the $[[7,1,3]]$ Steane code sit on opposite sides of a gate-set-versus-footprint tradeoff. The qRM code uses 15 physical qubits and admits a transversal logical $T$ gate, but it does not supply the full transversal Clifford group; the Steane code uses only seven physical qubits and admits transversal logical Clifford gates, but not a transversal logical $T$~\cite{paetznick2013universal,bravyi2012magic,steane1996multiple,bombin2007topological}. Converting between them lets the non-Clifford gate be executed in qRM while the resulting logical state is stored or processed in the smaller Steane block.

The prior work divides naturally by code pair and by switching mechanism. For the Steane--qRM pair itself, Anderson \emph{et al.} constructed bidirectional fault-tolerant conversion by representing the two codes as different gauge choices of a common subsystem code~\cite{anderson2014conversion}, and Quan \emph{et al.} reduced the required gauge measurements and extended the construction to adjacent Reed--Muller codes~\cite{quan2018conversion}. In this common-subsystem description the mechanism is \emph{gauge fixing}. Stabilizer code rewiring is a related but wider notion: it constructs measurement paths between stabilizer codes without requiring, by definition, a fixed common subsystem-code presentation~\cite{colladay2018rewiring}. Throughout, we use ``gauge fixing'' for the algebraic conversion implemented below and keep ``rewiring'' for the broader framework~\cite{bombin2015gauge,bombin2016dimensional,vuillot2019gauge}.

A separate line of work treats dimensional switching within the color code family. Kubica and Beverland combined transversal phase gates in higher-dimensional color codes with a Hadamard obtained by switching between color codes of different dimensions~\cite{kubica2015universal}; Butt \emph{et al.} then constructed deterministic and postselected flag-based circuits between distance-three two- and three-dimensional color codes~\cite{butt2024codeswitching}. Returning to the distance-three 7/15 setting, Daguerre and Kim proposed a two-block protocol that couples separately prepared qRM and Steane blocks through a transversal inter-block gate~\cite{daguerre2025code}. Heu{\ss}en and Hilder developed teleportation-based switching between the seven-qubit Steane code and the 15-qubit tetrahedral color code using logical auxiliary blocks and one-way transversal CNOT gates~\cite{heussen2025efficient}. Daguerre \emph{et al.} subsequently demonstrated qRM-to-Steane code switching and logical magic state preparation on a trapped-ion processor~\cite{daguerre2025experimental}.

Different choices of stabilizer generators describe the same codespace, but they need not give equivalent fault-tolerant implementations. Each measured generator comes with its own syndrome-extraction circuit, and the ancilla--data connectivity and gate ordering of that circuit control how physical faults propagate. Faults during a multiqubit check can leave correlated hook errors whose structure depends on the measurement sequence and the circuit layout. Flag ancillas add information about whether such a correlated fault has occurred, which the decoder and the acceptance rule can then use to correct or reject the event. The logical error rate of an implementation thus depends on the generator choice, the measurement ordering, the flag architecture, and the decoding strategy, and not on the underlying code alone. The dependence matters most when residual undetected Pauli errors are about to cross a non-Clifford operation. Flag-based schemes suppress dangerous correlated faults with a small number of additional ancillas~\cite{chao2018quantum,chamberland2018flag,tansuwannont2020flag,poor2026ultra}, while malignant-set analysis identifies the combinations of faults that dominate the leading logical error contribution~\cite{gottesman1998fault,aliferis2006quantum}.

The preceding protocols primarily reduce the number of measured gauge operators, optimize flag-assisted switching circuits, or transfer the logical state between separately encoded blocks. Our protocol instead treats the stabilizer-generator basis and the check ordering as fault-filtering design variables, and keeps the state inside one 15-qubit register throughout the qRM-to-Steane handoff. Eight qubits are measured in the $X$ basis and the remaining seven become the Steane block after a record-dependent software frame update, which removes the second encoded block together with its transversal block CNOT. Within the same shared-flag template, and without changing their span or gate count, the ten $Z$ checks are regrouped so that all accepted one- and two-fault logical contributions vanish under the matched error detection diagnostic. Exact low-order enumeration then gives a logical infidelity cubic in the physical error rate. An asymptotic order alone says nothing about the higher fault orders at a specific nonzero error rate, so we also construct a finite-$p$ certificate: a one-sided upper bound on the total conditional logical infidelity at $p=10^{-3}$, built from exact low-order enumeration, statistically bounded higher-order sampling, and a conservative tail bound. This certificate and the circuit resources are compared with the polynomial benchmark of Daguerre and Kim~\cite{daguerre2025code}. Three operational cases are kept distinct throughout: a destructive postselected diagnostic, standard error correction, and a retained unmeasured encoded output. The cubic advantage is established for the first case only.

Section~\ref{sec:preliminaries} collects the background on stabilizer and CSS codes, the qRM and Steane codes, subsystem codes and gauge fixing, code switching, and flagged syndrome extraction. The single-register gauge fix and the grouped flagged check circuits are defined in Sec.~\ref{sec:protocol-design}, and Sec.~\ref{sec:fault-order} sets up the accepted-fault sets and shows that no logical failure survives through fault order two. Section~\ref{sec:benchmarks} attaches circuit-level probabilities to those fault sets, builds the finite-$p$ certificate, and presents the logical-performance and resource benchmarks. Section~\ref{sec:discussion} returns to the stabilizer basis as a circuit-level design variable and to the limits of the destructive diagnostic, Sec.~\ref{sec:platforms} considers superconducting, neutral-atom, and trapped-ion implementations, and Sec.~\ref{sec:conclusion} concludes.

\section{Preliminaries}
\label{sec:preliminaries}

\subsection{Stabilizer and CSS codes}
\label{sec:stabilizer-css}

Let $\mathcal{P}_n$ denote the $n$-qubit Pauli group. For a subgroup $\mathcal A\subseteq\mathcal P_n$, its Pauli normalizer is $\mathcal N_{\mathcal P_n}(\mathcal A):=\{P\in\mathcal P_n:P\mathcal A P^\dagger=\mathcal A\}$. An $[[n,k,d]]$ stabilizer code is specified by an Abelian subgroup $\mathcal{S}\subset\mathcal{P}_n$ that does not contain $-I$. If $\mathcal{S}$ has $n-k$ independent generators, its simultaneous $+1$ eigenspace has dimension $2^k$ and encodes $k$ logical qubits. We write $\mathcal N(\mathcal S):=\mathcal N_{\mathcal P_n}(\mathcal S)$; logical Pauli operators are elements of $\mathcal N(\mathcal S)\setminus\mathcal S$, and two representatives that differ by an element of $\mathcal S$ act identically on the codespace. Here $\operatorname{wt}(P)$ denotes the number of nonidentity single-qubit tensor factors of $P$. The code distance is
\begin{equation}
d=\min_{P\in\mathcal{N}(\mathcal{S})\setminus\mathcal{S}}\operatorname{wt}(P),
\label{eq:stabilizer-distance}
\end{equation}
so a distance-three code corrects an arbitrary single-qubit Pauli error in the ideal error-correction setting.

For a binary vector $\bm{v}\in\GF^n$, define
\begin{equation}
X(\bm{v})=\prod_{j=1}^{n}X_j^{v_j},\qquad Z(\bm{v})=\prod_{j=1}^{n}Z_j^{v_j}.
\label{eq:binary-pauli}
\end{equation}

A Calderbank--Shor--Steane (CSS) code is defined by two binary linear
codes $C_X$ and $C_Z$ satisfying
\[
C_X\subseteq C_Z^\perp ,
\]
with stabilizer group
\begin{equation}
\mathcal{S}
=
\left\langle
X(\bm{x}),\bm{x}\in C_X;
\;
Z(\bm{z}),\bm{z}\in C_Z
\right\rangle .
\end{equation}
This separation allows bit-flip and phase errors to be detected independently through the corresponding $Z$- and $X$-type parity checks.

For any subset $S\subseteq\{1,\ldots,n\}$, we write $P_S=\prod_{j\in S}P_j$ for $P\in\{X,Z\}$; for example, $X_{1267}=X_1X_2X_6X_7$. Commas in such a subscript separate multi-digit labels only, so $Z_{4,5,11,12}=Z_4Z_5Z_{11}Z_{12}$.

The seven-qubit Steane code is the self-dual CSS code obtained from the classical $[7,4,3]$ Hamming code~\cite{steane1996multiple}. In the qubit labeling used throughout this work, a convenient generating set is
\begin{align}
\mathcal{S}_{\St}=\langle &X_{1267},X_{2347},X_{4567},\nonumber\\
&Z_{1267},Z_{2347},Z_{4567}\rangle,
\label{eq:prelim-steane-stabilizers}
\end{align}
with logical representatives
\begin{equation}
\logical{X}_{\St}=X_{123},\qquad \logical{Z}_{\St}=Z_{123}.
\label{eq:prelim-steane-logicals}
\end{equation}
The code has parameters $[[7,1,3]]$ and supports transversal logical Clifford operations, so it is a compact target for storing and processing a prepared magic state.

The 15-qubit quantum Reed--Muller code can be described by indexing the physical qubits by the nonzero vectors $\bm{v}\in\GF^4$. Let $\bm{r}_\mu\in\GF^{15}$ be the evaluation vector of the coordinate function $v_\mu$, and define
\begin{align}
C_X&=\operatorname{span}\{\bm{r}_1,\bm{r}_2,\bm{r}_3,\bm{r}_4\},\nonumber\\
C_Z&=\operatorname{span}\{\bm{1},\bm{r}_1,\bm{r}_2,\bm{r}_3,\bm{r}_4\}^{\perp}.
\label{eq:qrm-classical-spaces}
\end{align}
Its CSS stabilizer is
\begin{equation}
\mathcal{S}_{\qRM}=\langle X(\bm{x}):\bm{x}\in C_X,\; Z(\bm{z}):\bm{z}\in C_Z\rangle.
\label{eq:qrm-stabilizer-definition}
\end{equation}
This stabilizer has four independent $X$ generators and ten independent $Z$ generators and therefore defines an $[[15,1,3]]$ code. One may take $X(\bm{1})$ as a logical $X$ representative and any vector in $C_X^\perp\setminus C_Z$ as a logical $Z$ representative. With the labeling adopted below, we use the stabilizer-equivalent representatives
\begin{equation}
\logical{X}_{\qRM}=X_{1\cdots7},\qquad \logical{Z}_{\qRM}=Z_{123}.
\label{eq:prelim-qrm-logicals}
\end{equation}
The Reed--Muller structure permits a transversal non-Clifford logical $T$, with the physical $T/T^\dagger$ pattern fixed by the labeling convention~\cite{bravyi2012magic,paetznick2013universal}. The two gate structures are complementary, which suggests applying $T$ in the qRM code and then moving the logical state to the smaller Steane code.

\subsection{Subsystem codes and gauge fixing}
\label{sec:subsystem-gauge-prelim}

A subsystem code is a codespace $\mathcal C\subseteq(\mathbb C^2)^{\otimes n}$ that admits a Hilbert-space isomorphism
\begin{equation}
\mathcal{C}\cong\mathcal{H}_L\otimes\mathcal{H}_G.
\label{eq:subsystem-factorization}
\end{equation}
Here $\mathcal H_L$ is the protected logical subsystem, $\mathcal H_G$ is the gauge subsystem, and $\cong$ denotes an isomorphism rather than literal equality of subspaces. For an $[[n,k,d;r]]$ subsystem code, $\dim\mathcal H_L=2^k$ and $\dim\mathcal H_G=2^r$, so the code contains $k$ protected logical qubits and $r$ gauge qubits.

The code is specified by a generally non-Abelian gauge group $\mathcal G\subseteq\mathcal P_n$. For any subgroup $\mathcal A\subseteq\mathcal P_n$, define its Pauli centralizer by
\begin{equation}
C_{\mathcal P_n}(\mathcal A)
 :=\{P\in\mathcal P_n:PA=AP\ \text{for every }A\in\mathcal A\}.
\label{eq:pauli-centralizer}
\end{equation}
Up to the irrelevant Pauli phases, the stabilizer group of the subsystem code is
\begin{equation}
\mathcal S=\mathcal G\cap C_{\mathcal P_n}(\mathcal G).
\label{eq:gauge-center}
\end{equation}
Thus $\mathcal S$ is the commuting center of $\mathcal G$, has rank $n-k-r$, and fixes the full codespace $\mathcal C$. Elements of $\mathcal G$ may act nontrivially on $\mathcal H_G$ but act as the identity on $\mathcal H_L$. The protected distance is
\begin{equation}
d=\min_{P\in\mathcal N(\mathcal S)\setminus\mathcal G}\operatorname{wt}(P),
\label{eq:subsystem-distance}
\end{equation}
where $\mathcal N(\mathcal S)$ is the Pauli normalizer defined in Sec.~\ref{sec:stabilizer-css}.

Suppose $\mathcal G$ contains $r$ pairs of gauge-Pauli representatives $\{(X_{g_a},Z_{g_a})\}_{a=1}^{r}$ satisfying
\begin{equation}
X_{g_a}Z_{g_b}=(-1)^{\delta_{ab}}Z_{g_b}X_{g_a},
\qquad a,b\in\{1,\ldots,r\},
\label{eq:abstract-gauge-pairs}
\end{equation}
where $\delta_{ab}$ is the Kronecker delta; all of these operators commute with $\mathcal S$. Fixing the $Z$ gauge means adjoining a commuting set of signed operators $\{\pm Z_{g_a}\}$ to $\mathcal S$, whereas fixing the $X$ gauge means adjoining $\{\pm X_{g_a}\}$. This replacement of one complete commuting gauge set by another is called \emph{gauge fixing}. To convert from a $Z$-fixed sector to an $X$-fixed sector, one measures the target operators $X_{g_a}$ and uses the conjugate operators $Z_{g_a}$ as physical corrections or software frame updates for the measured signs~\cite{bombin2015gauge,bombin2016dimensional,vuillot2019gauge}. The measurement and frame update alter only $\mathcal H_G$; hence an ideal gauge fix preserves every state, including states entangled with a reference system, in the protected subsystem $\mathcal H_L$.

\begin{figure}[t]
\centering
  \resizebox{\columnwidth}{!}{%
    \input{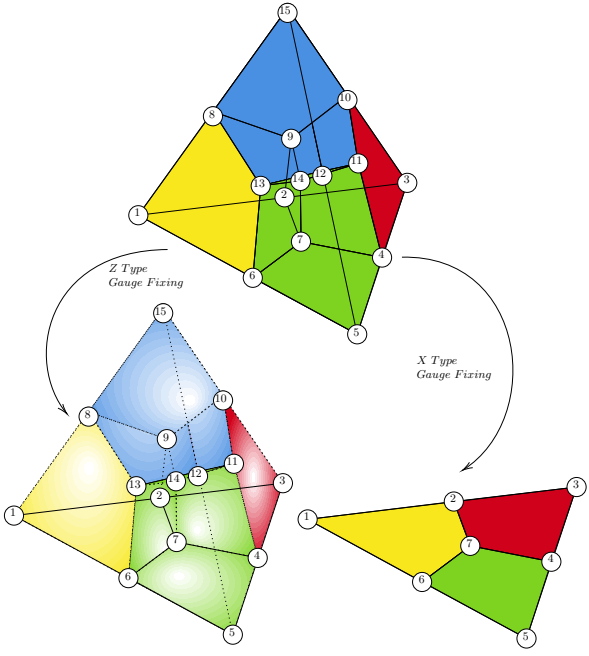}%
  }
\caption{ Geometric relation between the common stabilizer parent and its two gauge-fixed branches. The rank-11 common stabilizer group $\mathcal S_0$ defines the $[[15,4,3]]$ stabilizer parent shown at the top. Fixing the three $Z$-type gauge operators, $\mathcal S_{\qRM} =\langle\mathcal S_0,Z_{g_1},Z_{g_2},Z_{g_3}\rangle$, gives the $[[15,1,3]]$ qRM code shown at the lower left. Fixing the conjugate $X$-type gauge operators, $\mathcal S_{\mathrm{ERM}}  =\langle\mathcal S_0,X_{g_1},X_{g_2},X_{g_3}\rangle$, instead gives the 15-qubit extended-Steane sector. Measuring and discarding the auxiliary qubits $B=\{8,\ldots,15\}$ leaves the $[[7,1,3]]$ Steane code on $H=\{1,\ldots,7\}$, shown at the lower right. }
\label{fig:qrm-steane-gauge-fixing}
\end{figure}

For the qRM--Steane conversion used here, let
\begin{align}
h_1&=\{1,2,6,7\},&b_1&=\{8,9,13,14\},\nonumber\\
h_2&=\{2,3,4,7\},&b_2&=\{9,10,11,14\},\nonumber\\
h_3&=\{4,5,6,7\},&b_3&=\{11,12,13,14\},\nonumber\\
&&b_4&=\{8,9,10,11,12,13,14,15\},
\label{eq:main-parent-supports}
\end{align}
and define $c_a=h_a\mathbin{\dot\cup}b_a$ for $a=1,2,3$, where $\dot\cup$ denotes disjoint union, and $c_4=b_4$. The qRM and extended-Steane descriptions share the rank-11 stabilizer group
\begin{equation}
\begin{aligned}
\mathcal S_0=\langle
&X_{c_1},X_{c_2},X_{c_3},X_{c_4},
 Z_{c_1},Z_{c_2},Z_{c_3},Z_{c_4},\\
&Z_{h_1},Z_{h_2},Z_{h_3}\rangle .
\end{aligned}
\label{eq:main-common-center}
\end{equation}
Because $15-\operatorname{rank}(\mathcal S_0)=4$, this center defines a $[[15,4,3]]$ stabilizer parent. Designating one of its four encoded qubits as protected and the remaining three as gauge gives the $[[15,1,3;3]]$ subsystem code. A convenient set of gauge pairs is
\begin{align}
(X_{g_1},Z_{g_1})&=(X_{1267},Z_{4,7,11,14}),\nonumber\\
(X_{g_2},Z_{g_2})&=(X_{2347},Z_{4,5,11,12}),\nonumber\\
(X_{g_3},Z_{g_3})&=(X_{4567},Z_{2,7,9,14}).
\label{eq:main-gauge-pairs}
\end{align}
They obey Eq.~\eqref{eq:abstract-gauge-pairs}. The two rank-14 stabilizer choices are therefore
\begin{align}
\mathcal{S}_{\qRM}&=\langle\mathcal{S}_0,Z_{g_1},Z_{g_2},Z_{g_3}\rangle,\nonumber\\
\mathcal{S}_{\mathrm{ERM}}&=\langle\mathcal{S}_0,X_{g_1},X_{g_2},X_{g_3}\rangle.
\label{eq:prelim-two-gauges}
\end{align}
Here $\mathcal S_{\mathrm{ERM}}$ denotes the 15-qubit extended-Steane gauge. Its protected factor is the seven-qubit Steane code on the retained set $H=\{1,\ldots,7\}$, while the eight-qubit auxiliary factor on $B=\{8,\ldots,15\}$ is fixed independently of the protected logical state~\cite{anderson2014conversion,quan2018conversion}. Figure~\ref{fig:qrm-steane-gauge-fixing} summarizes the resulting geometric relation. Starting from the common $[[15,4,3]]$ stabilizer parent, fixing the three $Z$-type gauge operators yields the $[[15,1,3]]$ qRM code. Fixing the conjugate $X$-type gauge operators instead yields the extended-Steane sector; measuring and discarding the auxiliary qubits in $B$ then leaves the $[[7,1,3]]$ Steane code on $H$. Appendix~\ref{sec:codes} proves the corresponding factorization and branchwise conversion maps.

\begin{figure*}[t]
\centering
\input{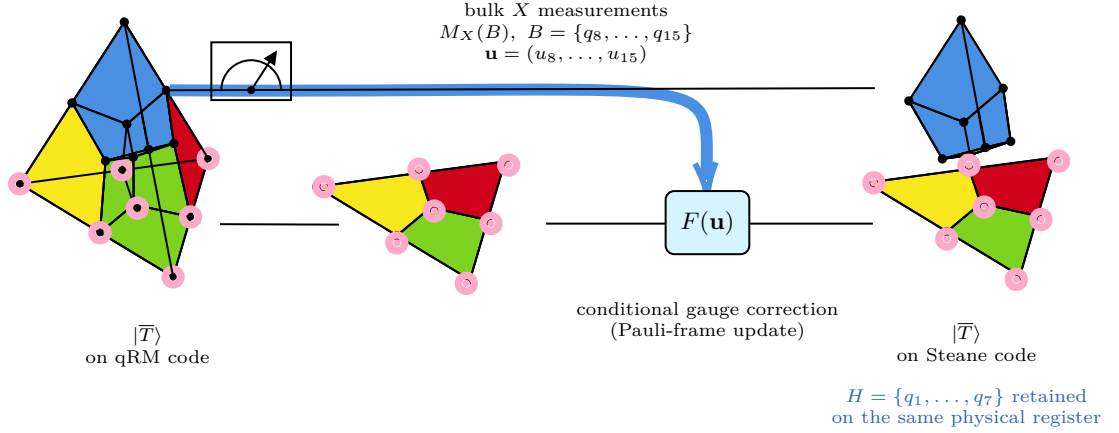}
\caption{ Comparison of two qRM-to-Steane handoff mechanisms for an encoded logical $T$ state. (a) In two-block teleportation-based code switching, a 15-qubit qRM block carrying the logical $T$ state is coupled to a separately prepared seven-qubit Steane block by a transversal inter-block CNOT. The qRM logical operator $\overline X$ is then measured destructively, and the measurement bit $m$ determines a conditional logical-frame update $\overline Z^{\,m}$ on the Steane block. The logical state is thereby transferred to a different physical register. (b) In the single-register gauge fix used in this work, the eight-qubit bulk $B=\{q_8,\ldots,q_{15}\}$ is measured destructively in the $X$ basis, while the seven-qubit hyperplane $H=\{q_1,\ldots,q_7\}$ is retained. The measurement record $\bm u=(u_8,\ldots,u_{15})$ determines the acceptance parity and a conditional gauge correction $F(\bm u)$, implemented as a Pauli frame update. The retained qubits in $H$ then constitute the live Steane output block. Highlighted vertices in panel (b) indicate the seven physical qubits retained throughout the in-place conversion. Unlike panel (a), panel (b) requires neither a second encoded Steane block nor an inter-block transversal CNOT. }
\label{fig:gauge-versus-teleportation}
\end{figure*}

\subsection{Code switching and measurement-based conversion}
\label{sec:code-switching-prelim}

Code switching is any fault-tolerant process that transfers an encoded logical state from a source code $A$ to a target code $B$ while preserving the logical information. Let $V_A$ and $V_B$ be encoding isometries for the same logical Hilbert space. For a measurement-based conversion with outcome record $\bm{m}$ and branch operator $K_{\bm{m}}$, ideal logical preservation means that
\begin{equation}
K_{\bm{m}}V_A=c_{\bm{m}}F_{\bm{m}}V_B
\label{eq:code-switching-kraus}
\end{equation}
for every accepted record, where $c_{\bm{m}}$ is a scalar and $F_{\bm{m}}$ is a known target-code frame update. Equation~\eqref{eq:code-switching-kraus} is stronger than checking a single input state: it shows that the complete logical channel, including arbitrary superpositions and entanglement with a reference system, is preserved branch by branch.

Gauge fixing is the restricted case of code switching in which the source and target are two complete gauge choices of one common subsystem code, as shown in Fig.~\ref{fig:gauge-versus-teleportation}(b). Teleportation-based switching instead uses separately encoded source and target blocks, an inter-block logical operation, and feed-forward, as in Fig.~\ref{fig:gauge-versus-teleportation}(a)~\cite{daguerre2025code,heussen2025efficient}. Stabilizer code rewiring is broader still: it describes sequences of Pauli measurements that map one stabilizer code to another and need not begin with a specified common gauge group~\cite{colladay2018rewiring}. Some rewiring or deformation paths do admit a gauge fixing interpretation~\cite{vuillot2019gauge}, but the two terms are not interchangeable.

The present construction is specifically a common-subsystem gauge fix. The source and target remain within one physical register; measuring the eight-qubit complement of a seven-point hyperplane fixes the required $X$-type gauge information, removes the bulk, and leaves the protected logical state on the seven retained qubits after a record-dependent Pauli frame update. Section~\ref{sec:gauge-fix} gives the explicit measurement relations and frame map.

\subsection{Flagged extraction, generator choice, and accepted logical faults}
\label{sec:flag-prelim}

\begin{figure}[t]
\centering
\resizebox{\columnwidth}{!}{\begin{tikzpicture}[x=0.72cm, y=0.72cm, font=\small, wire/.style={draw=black!78,line width=0.62pt}, gatewire/.style={draw=black!78,line width=0.62pt}, ctrl/.style={circle,fill=black!85,inner sep=0pt,minimum size=2.6mm}, targ/.style={
      circle,draw=black!85,line width=0.62pt,inner sep=0pt,minimum size=5.1mm,
      path picture={
        \draw[black!85,line width=0.55pt]
        (path picture bounding box.north)--(path picture bounding box.south);
        \draw[black!85,line width=0.55pt]
        (path picture bounding box.west)--(path picture bounding box.east);
      }
    }, prep/.style={draw=black!70,rounded corners=0.8pt,fill=white,inner sep=1.2pt}, meas/.style={draw=black!70,rounded corners=0.8pt,fill=white,inner sep=1.2pt,minimum width=5.0mm}, errburst/.style={
      starburst,starburst points=12,starburst point height=1.25mm,
      draw=orange!90!black,fill=orange!14,line width=0.65pt,
      inner sep=0pt,minimum size=4.7mm
    }, errbox/.style={
      draw=orange!90!black,fill=white,rounded corners=1.1pt,
      text=black,inner xsep=4pt,inner ysep=2pt
    }, zmark/.style={
      circle,draw=orange!90!black,text=orange!90!black,
      line width=0.75pt,inner sep=0pt,minimum size=5.2mm
    }, zfbox/.style={
      draw=orange!90!black,fill=white,rounded corners=1.0pt,
      text=black,inner xsep=3pt,inner ysep=2pt
    }, flagbox/.style={
      draw=teal!75!black,fill=white,rounded corners=1.0pt,
      text=teal!75!black,inner xsep=4pt,inner ysep=2.5pt
    }, prop/.style={
      orange!90!black,densely dashed,line width=0.75pt,
      -{Latex[length=1.8mm,width=1.2mm]}
    }, flagprop/.style={
      teal!75!black,densely dashed,line width=0.75pt,
      -{Latex[length=1.8mm,width=1.2mm]}
    }, paneltitle/.style={anchor=west,font=\sffamily\bfseries\large}, ordernote/.style={align=center,font=\normalsize}, resultnote/.style={align=center,font=\normalsize}, expl/.style={align=left,font=\normalsize,text width=5.8cm}]

  \node[paneltitle] at (-0.35,18.2)
    {(a) Moving the same faulty interaction later reduces the propagated support};
  \begin{scope}[shift={(0,13.1)}]
    \foreach \lab/\yy in {$d_1$/4,$d_2$/3,$d_3$/2,$d_4$/1,$s$/0}{
      \node[anchor=east] at (0,\yy) {\lab};
      \draw[wire] (0.38,\yy)--(7.10,\yy);
    }
    \node[prep] at (0.49,0) {$\lvert0\rangle$};
    \node[meas] at (6.82,0) {$Z$};

    \foreach \xx/\yy in {1.25/4,2.75/3,4.25/2,5.75/1}{
      \node[ctrl] at (\xx,\yy) {};
      \node[targ] at (\xx,0) {};
      \draw[gatewire] (\xx,\yy)--(\xx,-0.35);
    }

    \node[errburst] (aLerr) at (3.48,0) {};
    \node[errbox, anchor=north] (aLlab) at (3.26,-0.62)
      {$I_{d_2}Z_s$ component};
    \draw[orange!90!black, line width=0.6pt] (aLerr.south)--(aLlab.north);

    \draw[prop] (aLerr.north east) to[out=54, in=205] (4.17,1.84);
    \draw[prop] (aLerr.east)       to[out=78, in=221] (5.66,0.86);

    \node[zmark] at (6.35,2) {$Z$};
    \node[zmark] at (6.35,1) {$Z$};

    \node[ordernote] at (3.65,-1.48) {order $(d_1,d_2,d_3,d_4)$};
    \node[resultnote] at (3.65,-2.18)
      {$E_{\mathrm{data}}=Z_{d_3}Z_{d_4},\quad \operatorname{wt}=2$};
  \end{scope}
  \begin{scope}[shift={(10.25,13.1)}]
    \foreach \lab/\yy in {$d_1$/4,$d_2$/3,$d_3$/2,$d_4$/1,$s$/0}{
      \node[anchor=east] at (0,\yy) {\lab};
      \draw[wire] (0.38,\yy)--(7.10,\yy);
    }
    \node[prep] at (0.48,0) {$\lvert0\rangle$};
    \node[meas] at (6.82,0) {$Z$};

    \foreach \xx/\yy in {1.25/4,2.75/2,4.25/3,5.75/1}{
      \node[ctrl] at (\xx,\yy) {};
      \node[targ] at (\xx,0) {};
      \draw[gatewire] (\xx,\yy)--(\xx,-0.35);
    }

    \node[errburst] (aRerr) at (5.03,0) {};
    \node[errbox, anchor=north] (aRlab) at (4.76,-0.62)
      {$I_{d_2}Z_s$ component};
    \draw[orange!90!black, line width=0.6pt] (aRerr.south)--(aRlab.north);

    \draw[prop] (aRerr.north east) to[out=22, in=205] (5.66,0.86);
    \node[zmark] at (6.35,1) {$Z$};

    \node[ordernote] at (3.65,-1.48) {order $(d_1,d_3,d_2,d_4)$};
    \node[resultnote] at (3.65,-2.18)
      {$E_{\mathrm{data}}=Z_{d_4},\quad \operatorname{wt}=1$};
  \end{scope}
  \node[paneltitle] at (-0.35,10.35)
    {(b) At a fixed $Z_s$-insertion index, CNOT order changes the hook support};
  \begin{scope}[shift={(0,5.25)}]
    \foreach \lab/\yy in {$d_1$/4,$d_2$/3,$d_3$/2,$d_4$/1,$s$/0}{
      \node[anchor=east] at (0,\yy) {\lab};
      \draw[wire] (0.38,\yy)--(7.10,\yy);
    }
    \node[prep] at (0.49,0) {$\lvert0\rangle$};
    \node[meas] at (6.82,0) {$Z$};

    \foreach \xx/\yy in {1.25/4,2.75/3,4.25/2,5.75/1}{
      \node[ctrl] at (\xx,\yy) {};
      \node[targ] at (\xx,0) {};
      \draw[gatewire] (\xx,\yy)--(\xx,-0.35);
    }

    \node[errburst] (bLerr) at (3.48,0.01) {};
    \node[errbox, anchor=north] (bLlab) at (3.26,-0.62) {$Z_s$ component};
    \draw[orange!90!black, line width=0.6pt] (bLerr.south)--(bLlab.north);

    \draw[prop] (bLerr.north east) to[out=54, in=205] (4.17,1.84);
    \draw[prop] (bLerr.east)       to[out=78, in=221] (5.66,0.86);

    \node[zmark] at (6.35,2) {$Z$};
    \node[zmark] at (6.35,1) {$Z$};

    \node[ordernote] at (3.65,-1.48) {order $(d_1,d_2,d_3,d_4)$};
    \node[resultnote] at (3.65,-2.28)
      {$Z_s$ component after CNOT 2:\\[-1pt]$E_{\mathrm{data}}=Z_{d_3}Z_{d_4}$};
  \end{scope}
  \begin{scope}[shift={(10.25,5.25)}]
    \foreach \lab/\yy in {$d_1$/4,$d_2$/3,$d_3$/2,$d_4$/1,$s$/0}{
      \node[anchor=east] at (0,\yy) {\lab};
      \draw[wire] (0.38,\yy)--(7.10,\yy);
    }
    \node[prep] at (0.48,0) {$\lvert0\rangle$};
    \node[meas] at (6.82,0) {$Z$};

    \foreach \xx/\yy in {1.25/4,2.75/2,4.25/3,5.75/1}{
      \node[ctrl] at (\xx,\yy) {};
      \node[targ] at (\xx,0) {};
      \draw[gatewire] (\xx,\yy)--(\xx,-0.35);
    }

    \node[errburst] (bRerr) at (3.51,0.02) {};
    \node[errbox, anchor=north] (bRlab) at (3.26,-0.62) {$Z_s$ component};
    \draw[orange!90!black, line width=0.6pt] (bRerr.south)--(bRlab.north);

    \draw[prop] (bRerr.north east) to[out=60, in=205] (4.17,2.84);
    \draw[prop] (bRerr.east)       to[out=80, in=211] (5.66,0.86);

    \node[zmark] at (6.35,3) {$Z$};
    \node[zmark] at (6.35,1) {$Z$};

    \node[ordernote] at (3.65,-1.48) {order $(d_1,d_3,d_2,d_4)$};
    \node[resultnote] at (3.65,-2.28)
      {$Z_s$ component after CNOT 2:\\[-1pt]$E_{\mathrm{data}}=Z_{d_2}Z_{d_4}$};
  \end{scope}
  \begin{scope}[xshift=-5pt, yshift=-3pt]
    \node[paneltitle] at (-0.16,0.94)
              {(c) A flag ancilla detects the weight-two hook};
    \begin{scope}[shift={(0,-6.05)}]
              \foreach \lab/\yy in {$f$/5.6,$d_1$/4.3,$d_2$/3.2,$d_3$/2.1,$d_4$/1.0,$s$/0}{
                \node[anchor=east] at (0,\yy) {\lab};
                \draw[wire] (0.38,\yy)--(11.05,\yy);
              }
              \node[prep] at (0.72,5.6) {$\lvert+\rangle$};
              \node[meas] at (10.70,5.6) {$X$};
              \node[prep] at (0.72,0) {$\lvert0\rangle$};
              \node[meas] at (10.70,0) {$Z$};
          
              \node[ctrl] at (1.55,5.6) {};
              \node[targ] at (1.55,0) {};
              \draw[gatewire] (1.55,5.6)--(1.55,-0.35);
          
              \foreach \xx/\yy in {3.25/4.3,4.85/3.2,6.45/2.1,8.05/1.0}{
                \node[ctrl] at (\xx,\yy) {};
                \node[targ] at (\xx,0) {};
                \draw[gatewire] (\xx,\yy)--(\xx,-0.35);
              }
          
              \node[ctrl] at (9.75,5.6) {};
              \node[targ] (node1) at (9.75,0) {};
              \draw[gatewire] (9.75,5.6)--(9.75,-0.35);
          
              \node[errburst] (cErr) at (5.62,0) {};
              \node[errbox, anchor=north] (cLab) at (5.34,-0.72) {$Z_s$ component};
              \draw[orange!90!black, line width=0.6pt] (cErr.south)--(cLab.north);
          
              \draw[prop] (cErr.north east) to[out=66, in=205] (6.37,1.92);
              \draw[prop] (cErr.east)       to[out=70, in=211] (7.96,0.86);
          
              \node[zmark] at (10.55,2.1) {$Z$};
              \node[zmark] at (10.55,1.0) {$Z$};
          
              \coordinate (cRouteA) at (6.63,-0.52);
              \coordinate (cRouteB) at (7.12,-0.73);
              \draw[prop] (cErr.south east)--(cErr.south east)--(cErr.south east)
                to[out=325, in=217] (9.66,-0.08);
          
              \node[zfbox, anchor=west] (zf) at (11.25,0.72) {$Z_f$};
              \draw[prop] (9.88,0.12) to[out=18, in=190] (zf.west);
          
              \node[flagbox] (flagout) at (13.00,4.82) {$m_f^{(X)}=-1$};
              \draw[flagprop] (zf.north east) to[out=55, in=250] (flagout.south);
          
              \node[expl, anchor=west] at (0.54,-3.24)
                {$\mathrm{CNOT}_{f\to s}\,Z_s\,\mathrm{CNOT}_{f\to s}^{\dagger}=Z_fZ_s$.\\[5pt]
                Hence the weight-two hook $Z_{d_3}Z_{d_4}$ is accompanied by a nontrivial $X$-basis flag outcome.};
            \end{scope}
  \end{scope}

\end{tikzpicture}}
\caption{ Local hook-error propagation while measuring a weight-four $Z$ check with data-to-syndrome CNOTs. (a) The same $I_{d_2}Z_s$ fault component of the $d_2\rightarrow s$ interaction is compared under two CNOT schedules. In the order $(d_1,d_2,d_3,d_4)$, the residual $Z_s$ component crosses the two subsequent data CNOTs and produces the weight-two data error $Z_{d_3}Z_{d_4}$. Moving the same interaction to the third position in $(d_1,d_3,d_2,d_4)$ leaves only one subsequent data CNOT and reduces the propagated error to the weight-one operator $Z_{d_4}$. (b) When the $Z_s$ component is instead inserted after the second CNOT in both schedules, reordering changes the weight-two hook support from $Z_{d_3}Z_{d_4}$ to $Z_{d_2}Z_{d_4}$. (c) A schematic single-syndrome flag circuit uses opening and closing $f\rightarrow s$ CNOTs. These two couplings cancel in the ideal circuit, whereas a $Z_s$ component arising between them crosses the closing coupling as $Z_s\mapsto Z_fZ_s$ and therefore flips the final $X$-basis flag result to $m_f^{(X)}=-1$. The complete protocol uses shared-flag gadgets containing two or three syndrome rails; their exact CNOT schedules are given in Appendix~\ref{sec:gadgets}. }
\label{fig:hook-order}
\end{figure}
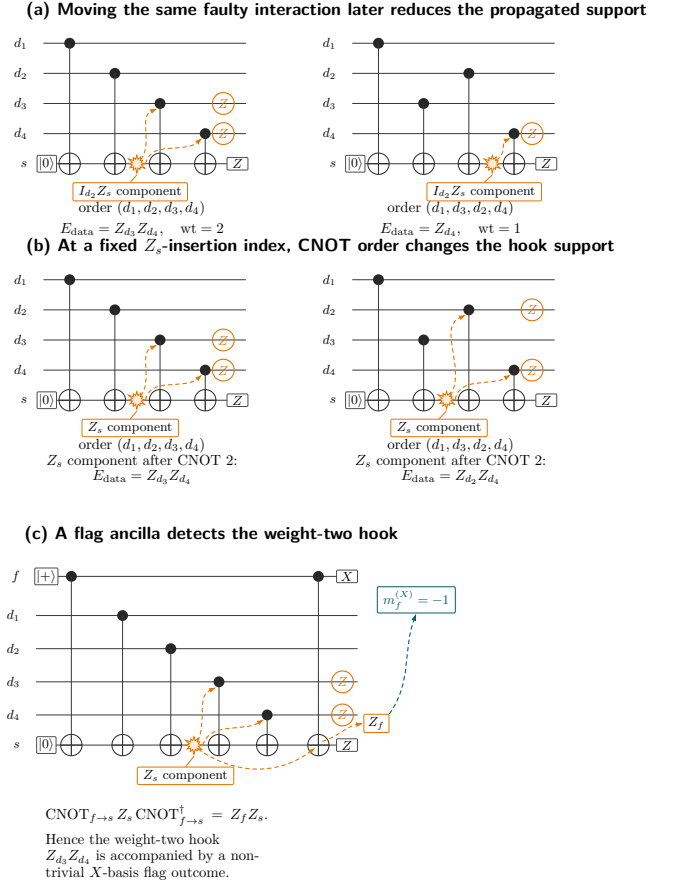

A stabilizer group does not have a unique generating set. If $\mathcal{S}=\langle s_1,\ldots,s_m\rangle$, then any invertible matrix $M\in\mathrm{GL}(m,\GF)$ defines another basis
\begin{equation}
\widetilde{s}_a=\prod_{i=1}^{m}s_i^{M_{ai}}\quad\Longrightarrow\quad\langle\widetilde{s}_1,\ldots,\widetilde{s}_m\rangle=\mathcal{S}.
\label{eq:generator-basis-change}
\end{equation}
The two bases define exactly the same ideal code. Their syndrome-extraction circuits, however, can have different fault-tolerance properties. The support of a generator determines which data qubits interact with the syndrome ancilla; the CNOT order determines how an ancilla fault propagates into a correlated data error; and the order in which the checks are executed determines which later outcomes can still reveal an error produced earlier. For a data-to-syndrome CNOT,
\begin{equation}
\mathrm{CNOT}_{d\rightarrow s}\,Z_s\, \mathrm{CNOT}_{d\rightarrow s}^{\dagger} = Z_dZ_s,
\label{eq:prelim-hook-propagation}
\end{equation}
so a $Z_s$ component is copied to every data control that interacts with the same syndrome target after the fault occurs. Figure~\ref{fig:hook-order} separates two consequences of the CNOT schedule: moving the same faulty interaction later can reduce the propagated data support, while reordering at a fixed insertion index can change the support of a hook without changing its weight. The figure also gives a schematic illustration of how opening and closing flag couplings record such propagation.

A flag qubit is coupled to the syndrome-extraction circuit so that a single fault capable of producing a potentially uncorrectable correlated data error also yields a nontrivial flag outcome or an otherwise distinguishable syndrome record~\cite{chao2018quantum,chamberland2018flag, tansuwannont2020flag}. The flag does not prevent the physical propagation shown in Fig.~\ref{fig:hook-order}. Instead, the opening and closing flag couplings convert a residual syndrome component into a detectable flag record. In the complete grouped gadgets used below, one flag is shared by two or three syndrome rails, so panel (c) illustrates only the local propagation mechanism rather than the full extraction topology.

The local flag criterion is therefore a one-fault detection condition, not by itself a proof of the final logical failure rate. That rate also depends on the complete generator basis, the CNOT and check orderings, the downstream bulk-parity and Steane detectors, and the rule by which the classical record is decoded or postselected.

More formally, let $F$ be a set of circuit faults, $D(F)$ the resulting syndrome-and-flag record, and $L(F)$ the induced logical Pauli class after the prescribed frame updates. A decoder maps $D(F)$ to a correction, while a postselection rule maps it to either acceptance or rejection. Neither rule alters the physical propagation that produced $D(F)$; they only sort the residual error into a correction, a rejection, or an accepted logical failure. Under pure error detection, an \emph{accepted logical fault} is a fault set satisfying
\begin{equation}
A[D(F)]=1,\qquad L(F)\ne I,
\label{eq:accepted-logical-fault}
\end{equation}
where $A$ is the acceptance indicator. Even with the abstract stabilizer code held fixed, the noisy preparation instrument is set by the generator basis together with the CNOT and check orderings, the flag topology, the decoder, and the acceptance conditions.

The syndrome and flag records are used differently in three operational tasks, so it matters which logical quantity is being bounded. In a \emph{destructive error detection diagnostic}, the output block is measured and nontrivial syndrome or flag outcomes are used as rejection conditions when inferring a logical observable. In standard error correction, the same records are decoder inputs: all correctable records are retained and mapped to corrections. For a \emph{live output fidelity}, the encoded block is not destructively measured and must be certified while remaining available for later computation. The flag record therefore plays a different role in each of the three tasks, and a bound proved for one of them does not transfer to the other two.

A second distinction separates an asymptotic order from a numerical guarantee. A low-noise expansion such as $I(p)=Cp^r+O(p^{r+1})$ identifies the leading fault order, but on its own it bounds nothing about the higher orders at a specified nonzero $p$. We use the term \emph{finite-$p$ certificate} for a statistically valid one-sided upper bound on the complete conditional logical infidelity at a fixed physical error rate, including controlled contributions from higher fault orders.

\section{Protocol design}
\label{sec:protocol-design}

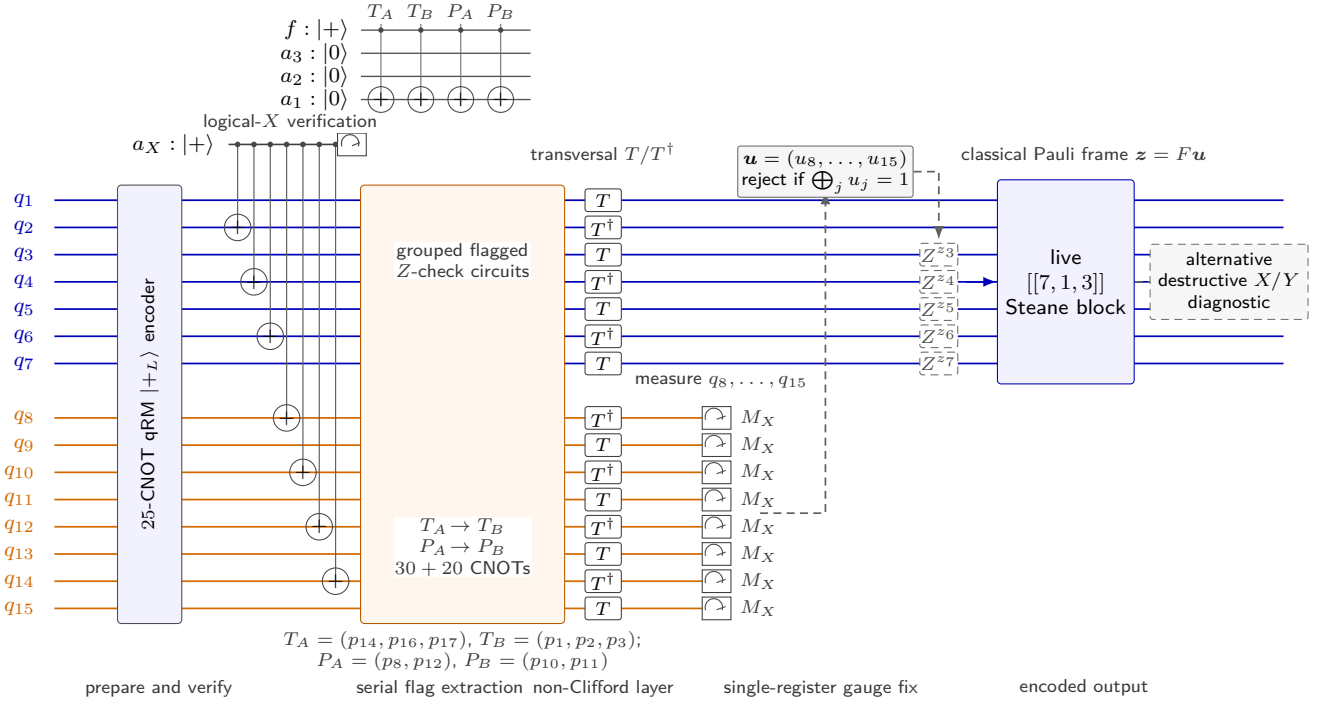
\begin{figure*}[t]
\centering
\begin{tikzpicture}[
  x=1.20cm,y=0.360cm,
  font=\sffamily\footnotesize,
  retained/.style={draw=blue!72!black,line width=0.65pt},
  measured/.style={draw=orange!82!black,line width=0.65pt},
  classical/.style={draw=black!62,densely dashed,line width=0.65pt,-{Latex[length=1.8mm]}},
  gate/.style={draw=black!75,fill=white,rounded corners=1pt,minimum width=4.8mm,
    minimum height=3.0mm,inner sep=0.25pt,font=\sffamily\scriptsize},
  prepblock/.style={draw=black!70,fill=blue!5,rounded corners=1.5pt,
    minimum width=8.5mm,minimum height=58mm,align=center,inner sep=1.4pt,
    font=\sffamily\scriptsize},
  checkblock/.style={draw=orange!72!black,fill=orange!7,rounded corners=1.5pt,
    minimum width=27mm,minimum height=58mm,align=center,inner sep=1.4pt,
    font=\sffamily\scriptsize},
  meter/.style={draw=black!75,fill=white,minimum width=3.8mm,minimum height=3.0mm,
    inner sep=0pt},
  note/.style={font=\sffamily\scriptsize,align=center,text=black!78,fill=white,inner sep=0.45pt}
]

\foreach \q/\yy in {1/15,2/14,3/13,4/12,5/11,6/10,7/9}{
  \node[anchor=east,text=blue!72!black] at (-0.12,\yy) {$q_{\q}$};
  \draw[retained] (0,\yy)--(13.55,\yy);
}
\foreach \q/\yy in {8/7,9/6,10/5,11/4,12/3,13/2,14/1,15/0}{
  \node[anchor=east,text=orange!82!black] at (-0.12,\yy) {$q_{\q}$};
  \draw[measured] (0,\yy)--(7.30,\yy);
}

\node[prepblock] (enc) at (1.05,7.5) {\rotatebox{90}{$25$-CNOT qRM $\ket{+_L}$ encoder}};

\draw[black!75,line width=0.6pt] (1.92,17.05)--(3.12,17.05);
\node[anchor=east] at (1.86,17.05) {$a_X:\ket{+}$};
\foreach \xx/\yy in {2.02/14,2.20/12,2.38/10,2.56/7,2.74/5,2.92/3,3.10/1}{
  \draw[black!62,line width=0.45pt] (\xx,17.05)--(\xx,\yy);
  \fill[black!72] (\xx,17.05) circle (1.05pt);
  \node[draw=black!72,circle,inner sep=0pt,minimum size=3.5mm,
    font=\sffamily\scriptsize] at (\xx,\yy) {$+$};
}
\node[meter] at (3.28,17.05) {};
\draw[black!70,line width=0.45pt] (3.17,16.98) arc[start angle=200,end angle=-20,radius=1.15mm];
\draw[black!70,line width=0.45pt] (3.28,17.00)--(3.37,17.12);
\node[note,anchor=south] at (2.60,17.48) {logical-$X$ verification};

\node[checkblock] (checks) at (4.50,7.5) {};
\foreach \name/\yy in {$a_1:|0\rangle$/18.7,$a_2:|0\rangle$/19.55,$a_3:|0\rangle$/20.4,$f:|+\rangle$/21.25}{
  \draw[black!68,line width=0.5pt] (3.38,\yy)--(5.25,\yy);
  \node[anchor=east] at (3.34,\yy) {\name};
}
\foreach \xx/\lab in {3.60/$T_A$,4.04/$T_B$,4.48/$P_A$,4.92/$P_B$}{
  \draw[black!58,line width=0.4pt] (\xx,18.45)--(\xx,21.48);
  \fill[black!72] (\xx,21.25) circle (1.0pt);
  \node[draw=black!72,circle,inner sep=0pt,minimum size=3.4mm,
    font=\sffamily\scriptsize] at (\xx,18.7) {$+$};
  \node[note,anchor=south] at (\xx,21.58) {\lab};
}
\node[note] at (4.50,12.8) {grouped flagged\\$Z$-check circuits};
\node[note] at (4.50,2.25) {$T_A\!\to T_B$\\$P_A\!\to P_B$\\$30+20$ CNOTs};
\node[note,anchor=north] at (4.50,-0.75) {$T_A=(p_{14},p_{16},p_{17})$, $T_B=(p_1,p_2,p_3)$;\\$P_A=(p_8,p_{12})$, $P_B=(p_{10},p_{11})$};

\foreach \q/\yy in {1/15,2/14,3/13,4/12,5/11,6/10,7/9,8/7,9/6,10/5,11/4,12/3,13/2,14/1,15/0}{
  \ifodd\q
    \node[gate] at (6.05,\yy) {$T$};
  \else
    \node[gate] at (6.05,\yy) {$T^\dagger$};
  \fi
}
\node[note,anchor=south] at (6.05,16.25) {transversal $T/T^\dagger$};

\foreach \q/\yy in {8/7,9/6,10/5,11/4,12/3,13/2,14/1,15/0}{
  \node[meter] at (7.30,\yy) {};
  \draw[black!70,line width=0.42pt] (7.18,\yy-0.07) arc[start angle=200,end angle=-20,radius=1.15mm];
  \draw[black!70,line width=0.42pt] (7.30,\yy-0.05)--(7.40,\yy+0.07);
  \node[note,anchor=west] at (7.54,\yy) {$M_X$};
}
\node[note,anchor=south] at (7.35,8.05) {measure $q_8,\ldots,q_{15}$};
\draw[classical] (7.78,3.5)--(8.50,3.5)--(8.50,15.35);
\node[draw=black!58,fill=gray!7,rounded corners=1pt,align=center,inner sep=1.4pt,
  font=\sffamily\scriptsize] (record) at (8.50,16.05)
  {$\bm u=(u_8,\ldots,u_{15})$\\reject if $\bigoplus_j u_j=1$};

\foreach \yy/\jj in {13/3,12/4,11/5,10/6,9/7}{
  \node[gate,densely dashed,draw=black!58,text=black!70] at (9.75,\yy) {$Z^{z_{\jj}}$};
}
\draw[classical] (record.east)--(9.75,16.05)--(9.75,13.55);
\node[note,anchor=south] at (11.35,16.42) {classical Pauli frame $\bm z=F\bm u$};

\node[draw=blue!72!black,fill=blue!5,rounded corners=1.4pt,align=center,
  minimum width=16mm,minimum height=27mm] (steane) at (11.15,12.0)
  {live\\$[[7,1,3]]$\\Steane block};
\draw[retained,-{Latex[length=1.8mm]}] (10.05,12.0)--(steane.west);
\draw[classical] (steane.east)--(12.55,12.0);
\node[draw=black!55,fill=gray!6,densely dashed,rounded corners=1pt,align=center,
  font=\sffamily\scriptsize] at (12.95,12.0)
  {alternative\\destructive $X/Y$\\diagnostic};

\node[note,anchor=north] at (1.15,-2.55) {prepare and verify};
\node[note,anchor=north] at (4.25,-2.55) {serial flag extraction};
\node[note,anchor=north] at (6.05,-2.55) {non-Clifford layer};
\node[note,anchor=north] at (8.45,-2.55) {single-register gauge fix};
\node[note,anchor=north] at (11.35,-2.55) {encoded output};

\end{tikzpicture}
\caption{Architecture and circuit flow of the single-register qRM-to-Steane preparation protocol. Blue wires ($q_1$--$q_7$) are retained; orange wires ($q_8$--$q_{15}$) are measured in $X$. The verifier ancilla controls seven CNOTs, and the four ancillas $a_1,a_2,a_3,f$ are reset and reused for two triple ($T_A,T_B$) and two pair ($P_A,P_B$) check circuits. Odd-labeled data qubits receive $T$ and even-labeled qubits receive $T^\dagger$. The record $\bm u$ supplies an even-parity detector and a software $Z$-frame update on the retained Steane face. Dashed arrows are classical; the dashed destructive $X/Y$ branch is an alternative characterization endpoint rather than part of the live output circuit. Exact check supports, CNOT orders, and frame equations are given in Appendices~\ref{sec:codes} and~\ref{sec:gadgets}.}
\label{fig:protocol}
\end{figure*}

Figure~\ref{fig:protocol} shows the architecture before the algebraic and circuit details are introduced. Against Daguerre--Kim we keep the 25-CNOT qRM encoder, the seven-CNOT logical-$X$ verifier, and the same four-gadget shared-flag resource envelope. Two coupled elements change: the rank-ten $Z$-generator basis together with its execution order, and the handoff itself, where a destructive bulk-$X$ gauge fix in the original 15-qubit register replaces the two-block teleportation. Reading the figure from left to right, the protocol (i) prepares and verifies $\ket{+_L}$ in qRM, (ii) extracts two flagged triples and two flagged pairs, (iii) applies transversal $T/T^\dagger$, (iv) measures the eight-qubit bulk and computes the parity detector and Pauli frame, and (v) retains the seven-qubit Steane block. The dashed branch after the live Steane block is an alternative destructive diagnostic used for the performance study; it is not an additional stage in a live output preparation shot.

\subsection{Single-register gauge fixing}
\label{sec:gauge-fix}

As reviewed in Sec.~\ref{sec:subsystem-gauge-prelim}, the qRM code and the 15-qubit extended-Steane description are two gauge fixings of the same $[[15,1,3;3]]$ subsystem code. We label the 15 physical qRM qubits by the nonzero vectors $\bm v=(v_1,v_2,v_3,v_4)\in\GF^4\setminus\{\bm 0\}$ and choose the seven-point hyperplane
\begin{equation}
H=\{q_1,\ldots,q_7\}=\{\bm v\neq\bm 0:v_4=0\}
\label{eq:retained-hyperplane}
\end{equation}
as the retained Steane block. Its complement is
\begin{equation}
B=\{q_8,\ldots,q_{15}\}=\{\bm v:v_4=1\},
\label{eq:measured-bulk}
\end{equation}
which is destructively measured in the $X$ basis.

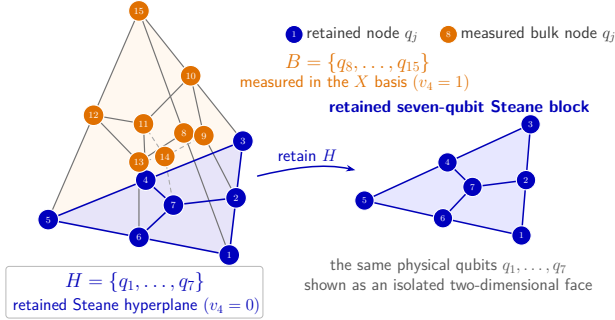
\begin{figure}[t]
\centering
\resizebox{\columnwidth}{!}{\definecolor{SteaneBlue}{HTML}{0000BD}
\definecolor{SteaneLine}{HTML}{0000B8}
\definecolor{BulkOrange}{HTML}{E07000}
\definecolor{BulkFace}{HTML}{FFF8F0}
\definecolor{LeftFace}{HTML}{E7E4F9}
\definecolor{RightFace}{HTML}{E7E7FF}
\definecolor{GraphGray}{HTML}{737373}
\definecolor{DashGray}{HTML}{A9A9A9}
\definecolor{TextGray}{HTML}{595959}
\definecolor{LegendGray}{HTML}{383838}
\definecolor{BoxGray}{HTML}{B8B8B8}

\begin{tikzpicture}[x=0.013725cm,y=-0.01372cm,line cap=round,line join=round]

  \coordinate (L1)  at (440,579);
  \coordinate (L2)  at (455,453);
  \coordinate (L3)  at (470,328);
  \coordinate (L4)  at (256,414);
  \coordinate (L5)  at (41,501);
  \coordinate (L6)  at (241,540);
  \coordinate (L7)  at (317,469);
  \coordinate (L8)  at (340,310);
  \coordinate (L9)  at (384,316);
  \coordinate (L10) at (355,184);
  \coordinate (L11) at (251,290);
  \coordinate (L12) at (141,271);
  \coordinate (L13) at (241,374);
  \coordinate (L14) at (298,362);
  \coordinate (L15) at (241,41);

  \coordinate (R1) at (1084,536);
  \coordinate (R2) at (1094,414);
  \coordinate (R3) at (1104,291);
  \coordinate (R4) at (921,375);
  \coordinate (R5) at (738,459);
  \coordinate (R6) at (911,498);
  \coordinate (R7) at (976,429);

  \fill[BulkFace] (L15)--(L3)--(L1)--(L5)--cycle;
  \fill[LeftFace] (L5)--(L3)--(L1)--cycle;
  \fill[RightFace] (R5)--(R3)--(R1)--cycle;

  \begin{scope}[draw=GraphGray,line width=1.02pt]
    \draw (L15)--(L12)--(L5);
    \draw (L15)--(L10)--(L3);
    \draw (L15)--(L8)--(L1);

    \draw (L8)--(L9)--(L10)--(L11)--(L12)--(L13)--cycle;

    \draw (L9)--(L2);
    \draw (L11)--(L4);
    \draw (L13)--(L6);
  \end{scope}

  \begin{scope}[draw=DashGray,line width=0.92pt,dash pattern=on 3.5pt off 3.4pt]
    \draw (L11)--(L14)--(L9);
    \draw (L13)--(L14);
    \draw (L14)--(L7);
  \end{scope}

  \begin{scope}[draw=SteaneLine,line width=1.35pt]
    \draw (L5)--(L4)--(L3);
    \draw (L3)--(L2)--(L1);
    \draw (L1)--(L6)--(L5);
    \draw (L4)--(L7)--(L2);
    \draw (L6)--(L7);
  \end{scope}

  \begin{scope}[draw=SteaneLine,line width=1.30pt]
    \draw (R5)--(R4)--(R3);
    \draw (R3)--(R2)--(R1);
    \draw (R1)--(R6)--(R5);
    \draw (R4)--(R7)--(R2);
    \draw (R6)--(R7);
  \end{scope}

  \tikzset{
    leftnode/.style={circle,draw=white,line width=1.0pt,minimum size=0.626cm,
      inner sep=0pt,fill=SteaneBlue,text=white,font=\fontsize{8.1}{8.1}\selectfont},
    bulknode/.style={leftnode,fill=BulkOrange},
    rightnode/.style={circle,draw=white,line width=0.95pt,minimum size=0.585cm,
      inner sep=0pt,fill=SteaneBlue,text=white,font=\fontsize{7.6}{7.6}\selectfont},
    legendnode/.style={circle,draw=white,line width=0.95pt,minimum size=0.547cm,
      inner sep=0pt,text=white,font=\fontsize{7.5}{7.5}\selectfont}
  }

  \foreach \n in {1,...,7}
    \node[leftnode] at (L\n) {\n};
  \foreach \n in {8,...,15}
    \node[bulknode] at (L\n) {\n};
  \foreach \n in {1,...,7}
    \node[rightnode] at (R\n) {\n};

  \node[legendnode, fill=SteaneBlue] at (582.89,93) {1};
  \node[anchor=west, inner sep=0pt, text=LegendGray, font=\Large\sffamily]
    at (610,93) {retained node $q_j$};
  \node[legendnode,fill=BulkOrange] at (923,92.82) {8};
  \node[anchor=west, inner sep=0pt, text=LegendGray, font=\Large\sffamily]
    at (950,93) {measured bulk node $q_j$};

  \node[anchor=base west, inner sep=0pt, text=BulkOrange, font=\LARGE, minimum width=166pt]
    at (510,166) {$B=\{q_8,\ldots,q_{15}\}$};
  \node[anchor=base west, inner sep=0pt, text=BulkOrange, font=\Large\sffamily, minimum width=236pt]
    at (414.01,209.55) {measured in the $X$ basis $(v_4=1)$};

  \node[anchor=center, text=SteaneLine, font=\Large\sffamily\bfseries]
    at (947.19,254.84) {retained seven-qubit Steane block};

  \draw[SteaneLine,line width=1.30pt,
        -{Stealth[length=8.0pt,width=5.7pt]}]
    (502.5,405) .. controls (577,387) and (647,376) .. (718,392);
  \node[anchor=center, text=SteaneLine, font=\Large\sffamily]
    at (612,360) {retain $H$};

  \node[anchor=north, align=center, text=TextGray, inner sep=0pt, font=\Large\sffamily]
    at (923,586) {the same physical qubits $q_1,\ldots,q_7$\\[-0.4pt]
                  shown as an isolated two-dimensional face};

  \draw[draw=BoxGray,line width=0.72pt,rounded corners=2.8pt]
    (-52.23,604.11) rectangle (514.25,717.98);
  \node[anchor=base, text=SteaneLine, font=\LARGE, inner sep=5.73pt]
    at (232.51,650.02) {$H=\{q_1,\ldots,q_7\}$};
  \node[anchor=base, text=SteaneLine, font=\Large\sffamily]
    at (235.77,700.81) {retained Steane hyperplane $(v_4=0)$};
\end{tikzpicture}}
\caption{ Three-dimensional geometric labeling of the 15 physical qubits of the $[[15,1,3]]$ qRM code and the seven-qubit Steane subsystem retained by the gauge fix. The blue nodes form the seven-point hyperplane $H=\{q_1,\ldots,q_7\}=\{\bm v\neq\bm0:v_4=0\}$, which remains as the live Steane block. The orange nodes form the complementary set $B=\{q_8,\ldots,q_{15}\}=\{\bm v:v_4=1\}$ and are measured destructively in the $X$ basis. The right-hand diagram isolates the same physical qubits $q_1,\ldots,q_7$ to display the retained Steane geometry; it does not denote a separate register or an inter-block transfer. Lines and shaded regions indicate the geometric embedding and the highlighted retained face, rather than physical circuit couplings. }
\label{fig:qrm-hyperplane}
\end{figure}

Figure~\ref{fig:qrm-hyperplane} displays this partition in a three-dimensional tetrahedral representation of the 15-qubit qRM geometry. The seven qubits in $H$ are highlighted as the blue Steane face, while the remaining eight qubits in $B$ are shown in orange. The isolated face on the right repeats the same physical qubits $q_1,\ldots,q_7$ to make the retained Steane geometry explicit.

The geometric extraction in Fig.~\ref{fig:qrm-hyperplane} should not be interpreted as teleportation from a 15-qubit block to a separately prepared seven-qubit block. The protected logical information is already supported on the retained hyperplane $H$. Measuring the complementary set $B$ fixes the remaining gauge degrees of freedom, disentangles and removes the bulk, and leaves an encoded state on the original physical qubits $q_1,\ldots,q_7$ after a record-dependent Pauli frame update.

The eight physical readouts are therefore not eight independent gauge measurements. Together they refine the three commuting $X$-type gauge outcomes, remove the eight-qubit bulk, and supply one additional parity detector. This single-register handoff requires neither a second encoded block nor an inter-block transversal CNOT.

Let $\bm u=(u_8,\ldots,u_{15})\in\GF^8$ denote the bulk measurement record, where $u_j=0$ for the $+1$ eigenvalue of $X_j$, $u_j=1$ for the $-1$ eigenvalue, and $\oplus$ denotes addition modulo two. The record determines three frame bits associated with the Steane $X$ stabilizers in Eq.~\eqref{eq:prelim-steane-stabilizers}. It also determines the linear bulk-parity map $g:\GF^8\rightarrow\GF$:
\begin{align}
s_1(\bm u)&=u_8\oplus u_9\oplus u_{13}\oplus u_{14},\nonumber\\
s_2(\bm u)&=u_9\oplus u_{10}\oplus u_{11}\oplus u_{14},\nonumber\\
s_3(\bm u)&=u_{11}\oplus u_{12}\oplus u_{13}\oplus u_{14},\nonumber\\
g(\bm{u})&:=\bigoplus_{j=8}^{15}u_j.
\label{eq:bulk-relations}
\end{align}
Thus $g(\bm u)=0$ labels even-parity bulk records and $g(\bm u)=1$ labels odd-parity records.
For a measurement branch $\bm{u}$, the three Steane $X$-check eigenvalues are $(-1)^{s_1(\bm u)}$, $(-1)^{s_2(\bm u)}$, and $(-1)^{s_3(\bm u)}$, respectively, while an ideal conversion satisfies $g(\bm{u})=0$. The record-dependent Pauli frame
\begin{equation}
F(\bm{u})=Z_{47}^{s_1(\bm u)}Z_{45}^{s_2(\bm u)}Z_{27}^{s_3(\bm u)}
\label{eq:bulk-frame}
\end{equation}
maps every even-parity branch to the standard Steane frame. Because this update is tracked in software, the three bits $s_1(\bm u),s_2(\bm u),s_3(\bm u)$ are corrected rather than postselected; only a nontrivial parity value $g(\bm{u})=1$ is rejected.

The protected logical information remains entirely on $H$. By Eqs.~\eqref{eq:prelim-steane-logicals} and~\eqref{eq:prelim-qrm-logicals}, the qRM and Steane logical $Z$ representatives coincide on the retained qubits, while their logical $X$ representatives differ only by the Steane stabilizer $X_{4567}$. Measuring $B$ therefore neither reveals nor removes the logical amplitudes. The eight measurement outcomes decompose into one parity detector, three gauge-frame bits, and four branch labels that act trivially on the retained logical subsystem. Appendix~\ref{sec:bulk} gives the explicit codespace factorization and proves the conversion identity branch by branch.

\subsection{Grouped flagged source checks}
\label{sec:grouped-flags}
 The Daguerre--Kim protocol prepares the qRM source block with a 25-CNOT $\ket{+_L}$ encoder followed by a seven-CNOT logical-$X$ verifier~\cite{daguerre2025code}. To suppress errors that would propagate through the transversal $T/T^\dagger$ layer, it then measures a rank-ten basis of the qRM $Z$-stabilizer space and postselects on trivial syndrome and flag outcomes. Instead of measuring the ten checks in ten separately flagged circuits, Daguerre and Kim adapt simultaneous shared-flag extraction and organize them into two two-check gadgets and two three-check gadgets. In the plaquette labeling of Ref.~\cite{daguerre2025code}, their basis is $\{p_1,p_2,p_3,p_7,p_8,p_9,p_{13},p_{16},p_{17},p_{18}\}$, grouped as the pairs $(p_{13},p_9)$ and $(p_7,p_8)$ and the triples $(p_1,p_2,p_3)$ and $(p_{16},p_{17},p_{18})$. This reduces the source-verification requirement to three syndrome ancillas and one shared flag ancilla, which can be reset and reused across the four gadgets. After the qRM block passes verification and receives transversal $T/T^\dagger$, their code switching stage prepares a separate Steane $\ket{0_L}$ block, applies a transversal block CNOT, destructively measures all 15 qRM qubits in the $X$ basis, postselects on the inferred qRM $X$ syndrome, and applies a record-dependent logical-$Z$ correction to the Steane block. The role of their grouped flag circuits is low-overhead qRM source verification; the logical handoff itself is still a two-block one-bit teleportation.

Our protocol retains the same 25-CNOT qRM encoder, seven-CNOT logical-$X$ verifier, ten measured $Z$ checks, two-triple/two-pair shared-flag resource template, four reusable ancillas, and 50-CNOT source-syndrome-extraction cost. We therefore do not obtain the improvement by adding another verification round or enlarging the local flag gadgets. Instead, we modify two coupled parts of the circuit. First, Sec.~\ref{sec:gauge-fix} replaces the two-block teleportation by a single-register bulk-$X$ gauge fix. Second, within the unchanged shared-flag template, we replace the Daguerre--Kim rank-ten generator basis and grouping by a basis and order chosen for the detector structure of this single-register handoff.

For self-containment, the ten $Z$-type plaquette checks used here are
\begin{alignat}{3}
p_1&=Z_{1267},&\quad p_2&=Z_{2347},&\quad p_3&=Z_{4567},\nonumber\\
p_8&=Z_{4,5,11,12},& p_{10}&=Z_{6,7,13,14},& p_{11}&=Z_{2,7,9,14},\nonumber\\
p_{12}&=Z_{4,7,11,14},& p_{14}&=Z_{8,9,10,15},& p_{16}&=Z_{8,9,13,14},\nonumber\\
&&p_{17}&=Z_{9,10,11,14}.&&
\label{eq:selected-plaquette-supports}
\end{alignat}
Three syndrome ancillas prepared in $\ket{0}$ and one flag ancilla prepared in $\ket{+}$ are reset and reused in the fixed order
\begin{equation}
(p_{14},p_{16},p_{17})\longrightarrow(p_1,p_2,p_3)\longrightarrow(p_8,p_{12})\longrightarrow(p_{10},p_{11}).
\label{eq:grouping}
\end{equation}
The first two groups use three-check gadgets and the last two use two-check gadgets. The selected plaquettes are distinct, have binary rank ten, and span the same complete qRM $Z$-stabilizer space as the earlier basis. The ideal qRM state, the number of extracted syndromes, and the local ancilla resources are unchanged; only the physical extraction instrument differs. The supports of the selected generators and their placement in the four gadgets alter which correlated hook errors are produced and which syndrome--flag signatures remain available to detect them.

The simultaneous shared-flag topology originates in Ref.~\cite{chao2018quantum}, was adapted to qRM verification in Ref.~\cite{daguerre2025code}, and satisfies the general flag criterion of Refs.~\cite{chamberland2018flag,tansuwannont2020flag}. If $U_2$ is a two-check gadget for $A=\{a,b,u_1,u_2\}$ and $B=\{a,b,v_1,v_2\}$, backward propagation gives
\begin{equation}
U_2^\dagger(Z_{s_1},Z_{s_2},X_f)U_2=(Z_AZ_{s_1},Z_BZ_{s_1}Z_{s_2},X_f).
\label{eq:pair-heisenberg}
\end{equation}
The analogous three-check circuit returns $Z_A$, $Z_B$, and $Z_C$ from three syndrome ancillas and restores $X_f$. A single flag can therefore protect two or three parity measurements without mixing their ideal outcomes. The propagation rules $Z_s\mapsto Z_dZ_s$ through a data-to-syndrome CNOT and $Z_s\mapsto Z_fZ_s$ through a flag-to-syndrome CNOT explain the local protection: the same syndrome-ancilla $Z$ component that creates a multi-data hook also flips the final $X_f$ result.

The two constructions therefore differ not in the flag principle but in the joint design of an equivalent stabilizer-generator basis, its grouping and execution order, and the downstream conversion circuit. In the earlier protocol, the grouped checks verify the qRM source before a two-block teleportation. In our protocol, the generator basis is evaluated together with the complete detector set that follows it: the remaining source groups, the bulk-parity detector generated by the single-register gauge fix, and the final Steane error detection checks. Such later detectors can expose hook patterns that the local flag record alone does not distinguish. The standard flag criterion constrains dangerous errors from a single fault inside one gadget; it is necessary here, but it does not imply the suppression of accepted two-fault events. Exhaustive propagation through the transversal $T/T^\dagger$ layer, the bulk gauge fix, the record-dependent frame, and the destructive error detection diagnostic shows that every accepted fault set of size at most two acts trivially on the logical qubit, which establishes $B_1=B_2=0$ for this generator basis and order. Exact CNOT lists, the geometric conditions for the pair and triple gadgets, and the complete fault-order test are given in Appendix~\ref{sec:gadgets}.

An attempt is retained only if the logical-$X$ verifier, all ten source syndromes, and all four source flags return their ideal values. Transversal $T$ on odd labels and $T^\dagger$ on even labels then implements logical $T$ on the qRM block, after which the eight bulk measurements complete the single-register gauge fix. The generator regrouping leaves the source-verification resource count unchanged, whereas the single-register handoff removes the separately encoded Steane block, its transversal block CNOT, and the destructive 15-qubit qRM logical-$X$ readout. Under the common counting convention summarized in Table~\ref{tab:resources}, the full circuit is thereby reduced from 26 to 19 qubits, from 100 to 82 CNOTs, from 31 to 23 measurements, and from component depth 81 to 70, with 15 physical $T/T^\dagger$ gates in both cases. The seven remaining qubits form an unmeasured encoded output before any destructive characterization. In Fig.~\ref{fig:protocol}, the solid endpoint denotes that live Steane block, whereas the dashed $X/Y$ branch denotes a separate destructive-characterization experiment used to define the matched diagnostic below; the two endpoints are not executed in the same shot.

\section{Fault-order suppression}
\label{sec:fault-order}

Accepted logical errors first appear at third order. This is a combinatorial property of the circuit and is independent of any numerical value assigned to the physical error probability. We define the fault alphabet here and postpone the probability model to Sec.~\ref{sec:benchmarks}. A fault set $F$ is a finite set of faulty circuit locations, with one allowed preparation or live-measurement flip at a preparation or measurement location, a nonidentity two-qubit Pauli after a CNOT, or a nonidentity one-qubit Pauli after a physical $T/T^\dagger$. The seven final Steane tomography measurements are not locations in the live preparation circuit. Let $|F|$ denote the number of faulty locations, $D(F)$ the complete source-syndrome, source-flag, bulk-parity, and final-Steane detector record, and $L(F)$ the induced logical Pauli class after the frame update in Eq.~\eqref{eq:bulk-frame}.

For fault order $w$, define the set of malignant accepted patterns by
\begin{equation}
\mathfrak M_w:=\{F:|F|=w,\ A[D(F)]=1,\ L(F)\ne I\},
\label{eq:malignant-accepted-set}
\end{equation}
where $A$ is the acceptance indicator from Eq.~\eqref{eq:accepted-logical-fault}. Because a Pauli error crossing $T/T^\dagger$ may acquire a local $S$ or $S^\dagger$ factor, the enumeration propagates the complete physical Clifford frame rather than only a Pauli label, and the four stabilizer branches of $\ket{T}$ are combined before conditional normalization. Complete stabilizer-tableau propagation~\cite{aaronson2004simulation,gidney2021stim} over the one- and two-fault subsets gives
\begin{equation}
\mathfrak M_1=\mathfrak M_2=\varnothing.
\label{eq:no-malignant-through-two}
\end{equation}
Equivalently, if $B_w$ denotes the total model weight of accepted logical patterns at order $w$, then $B_1=B_2=0$. Separate enumerations of the two-before-$T$, one-on-each-side, and two-after-$T$ partitions all satisfy Eq.~\eqref{eq:no-malignant-through-two}. The conclusion is therefore independent of any fitted curve or chosen value of $p$: for the matched destructive-ED diagnostic, the reported generator basis and order empty the first- and second-order accepted-logical fault sets completely.

\section{Numerical performance and resource benchmarks}
\label{sec:benchmarks}

Probabilities are now assigned to the fault alphabet of the previous section. Figure~\ref{fig:benchmark} compares conditional logical infidelity and acceptance over the simulated error-rate sweep. Table~\ref{tab:performance} isolates the model-matched comparison at $p=10^{-3}$, while Table~\ref{tab:resources} identifies which circuit stages produce the reductions to 19 qubits and 82 CNOTs.

\subsection{Circuit-level noise model and finite-\texorpdfstring{$p$}{p} certificate}
\label{sec:numerical-noise-model}

We adopt the uniform circuit-level model used for the closest published comparison~\cite{daguerre2025code}: preparation and live-measurement flips occur independently with probability $p$, every CNOT is followed by two-qubit depolarizing noise of strength $p$, every physical $T/T^\dagger$ is followed by one-qubit depolarizing noise of strength $p$, and idle errors are absent. The seven final Steane measurements are ideal destructive tomography operations in this convention. They determine a logical observable but are neither noisy locations nor live circuit resources.

Applying these probabilities to the exact fault-order enumeration gives
\begin{equation}
 I_{\rm ED}(p)=210.18p^3+O(p^4),
 \qquad B_1=B_2=0.
\label{eq:ed-series}
\end{equation}
The first two coefficients vanish because the sets in Eq.~\eqref{eq:no-malignant-through-two} are empty; no cancellation in a numerical fit is involved.

At $p=10^{-3}$, fault orders zero through three are evaluated exactly. Orders four, five, and six use $10^6$ uniformly weighted fixed-fault samples per order. One-sided Clopper--Pearson limits~\cite{clopper1934binomial} are Bonferroni-corrected, and every event of order seven or higher is conservatively treated as a failure. Dividing the resulting unnormalized loss bound by an acceptance lower bound gives
\begin{align}
 I_{\rm ED}(10^{-3})&\le 2.19\times10^{-7}
 \quad (99.9\%\ \text{joint confidence}),\nonumber\\
 P_{\rm ED}(10^{-3})&=86.90\%.
\label{eq:finite-certificate}
\end{align}
The exact interval for $P_{\rm ED}$ and the complete finite-$p$ construction are summarized in Appendix~\ref{sec:dkm-exact-finite}.

The quoted acceptance is the probability of the complete error detection event, not of the bulk parity alone. If $\bm d_{\rm src}\in\GF^{15}$ denotes the logical-$X$ verifier bit, ten source-syndrome bits, and four source-flag bits; $g(\bm u)\in\GF$ is the bulk detector defined in Eq.~\eqref{eq:bulk-relations}; and $\bm t\in\GF^3$ is the final Steane syndrome, then
\begin{align}
 P_{\rm ED}(p)&=\Pr_p[\bm d_{\rm src}=\bm0,\ g(\bm u)=0,\ \bm t=\bm0],\nonumber\\
 P_{\rm rej}(p)&=1-P_{\rm ED}(p).
\label{eq:acceptance-event}
\end{align}
These conditions are correlated, so the 13.10\% total rejection cannot be assigned to the odd-bulk event by subtraction. In particular, the three random gauge signs in Eq.~\eqref{eq:bulk-relations} are frame-corrected and are not rejection bits.

\subsection{Logical-performance benchmark}
\label{sec:logical-benchmark}

\begin{figure*}[t]
\centering
\includegraphics[width=0.98\textwidth]{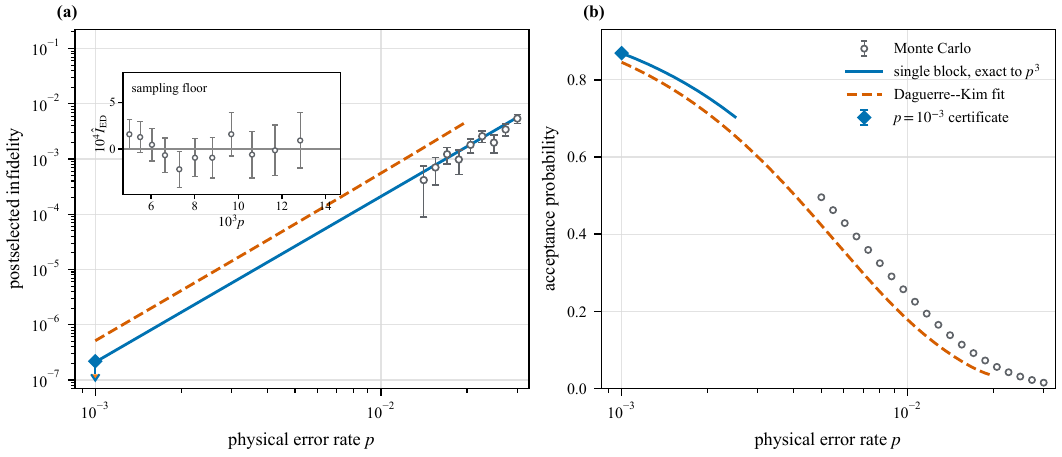}
\caption{Matched ideal-readout ED benchmark. (a) The diamond and downward arrow at $p=10^{-3}$ show the 99.9\% upper bound of Eq.~\eqref{eq:finite-certificate}; it is not a Monte Carlo point. Open circles show direct four-branch Monte Carlo estimates only where the complete one-standard-error interval is positive and can therefore be displayed on a logarithmic axis. The inset retains every remaining estimate on a linear axis, including negative values of the signed estimator. No point is clipped or horizontally displaced. (b) All 20 Monte Carlo acceptance points are shown. The blue curve is the exact expansion through $p^3$ and is drawn only in its low-$p$ range; the orange dashed curve is the published Daguerre--Kim polynomial fit and is stopped at $p=2\times10^{-2}$. The Monte Carlo abscissae form a logarithmic grid from $5\times10^{-3}$ to $3\times10^{-2}$, with $10^7$ shots for each of four stabilizer branches and two bases at every point.}
\label{fig:benchmark}
\end{figure*}

Figure~\ref{fig:benchmark} is the graphical summary of the logical-performance calculation. Panel (a) compares the conditional destructive-ED infidelity with the Daguerre--Kim polynomial and distinguishes the certified point at $p=10^{-3}$ from direct Monte Carlo estimates. Panel (b) plots the corresponding complete acceptance probability. Table~\ref{tab:performance} then reports the values at $p=10^{-3}$ and separates the model-matched Daguerre--Kim comparison from broader literature context whose noise and output conventions differ.

\begin{table}[t]
\caption{Performance at $p=10^{-3}$. The first two rows are model matched under ideal destructive error detection with no idle faults; the remaining rows provide literature context with different noise and output conventions.}
\label{tab:performance}
\begingroup
\scriptsize
\setlength{\tabcolsep}{2.0pt}
\renewcommand{\arraystretch}{1.08}
\begin{ruledtabular}
\begin{tabular}{@{}lcccc@{}}
Protocol & Output error & Acceptance & \shortstack{Final\\code} & $d$\\
\hline
Ours
 & \shortstack{$\le2.19\!\times\!10^{-7}$}
 & $86.90\%$
 & Steane
 & 3\\
Daguerre--Kim~\cite{daguerre2025code}
 & \shortstack{$5.13\!\times\!10^{-7}$}
 & $84.55\%$
 & Steane
 & 3\\
\hline
Gidney \emph{et al.}~\cite{gidney2024magic}
 & \shortstack{$\sim6\!\times\!10^{-7}$}
 & $\sim65\%$
 & \shortstack{Steane/\\color}
 & 3\\
Itogawa \emph{et al.}~\cite{itogawa2025even}
 & $\sim1.0\!\times\!10^{-4}$
 & $\sim70\%$
 & surface
 & 3
\end{tabular}
\end{ruledtabular}
\endgroup
\end{table}

The upper bound in Eq.~\eqref{eq:finite-certificate} lies below the value obtained by evaluating the published Daguerre--Kim fit at the same $p$. The two numbers have different statistical meanings: ours is a one-sided certificate, whereas the reference value is a low-$p$ extrapolation. We use the comparison only as a model-matched benchmark. The Gidney and Itogawa rows in Table~\ref{tab:performance} retain the definitions and precision of their source papers and are included only to show broader published context; their idle and processing conventions differ from the first two rows.

The plotted sweep serves as a consistency check; it is not the source of the $10^{-3}$ claim. Direct tomography cannot resolve a cubic signal there at practical shot counts, and the reported low-$p$ value comes from the exact enumeration and the finite-$p$ certificate. The logarithmic abscissae are shown at their actual values, and the branch counts determine the displayed estimates and standard errors.

\subsection{Resource benchmark}
\label{sec:resource-benchmark}

Table~\ref{tab:resources} decomposes the resource comparison stage by stage. The source encoder and the four grouped extraction gadgets are held fixed on purpose, which lets the savings be attributed to the handoff: the single-register construction drops the separately prepared Steane block and the transversal block CNOT, and replaces the 15-qubit qRM logical-$X$ readout by eight bulk-$X$ measurements.

\begin{table*}[t]
\caption{Stage-by-stage resource comparison with the same conventions as Table~VII of Ref.~\cite{daguerre2025code}. Here $V$ is the active-qubit-time sum, $n_2$ is the CNOT count, $M$ counts live destructive measurements, and $d$ is component depth. Seven-qubit tomography, routing, waiting idles, reset, and leakage are excluded in both columns. Peak width with ancilla reuse is 19 qubits here and 26 qubits for Daguerre--Kim.}
\label{tab:resources}
\begingroup
\scriptsize
\setlength{\tabcolsep}{3.8pt}
\begin{ruledtabular}
\begin{tabular}{lrrrr|rrrr}
&\multicolumn{4}{c|}{Our protocol}&\multicolumn{4}{c}{Daguerre--Kim~\cite{daguerre2025code}}\\
Stage & $V$ & $n_2$ & $M$ & $d$ & $V$ & $n_2$ & $M$ & $d$\\
\hline
Steane $\ket{0_L}$ preparation &--&--&--&-- &75&11&1&10\\
qRM $\ket{+_L}$ preparation &309&32&1&20 &309&32&1&20\\
two pair extractions &360&20&6&20 &360&20&6&20\\
two triple extractions &532&30&8&28 &532&30&8&28\\
transversal $T/T^\dagger$ &15&0&0&1 &15&0&0&1\\
block CNOT &--&--&--&-- &22&7&0&1\\
bulk-$X$ handoff &15&0&8&1 &--&--&--&--\\
qRM logical-$X$ measurement &--&--&--&-- &22&0&15&1\\
\hline
total &1231&82&23&70 &1335&100&31&81
\end{tabular}
\end{ruledtabular}
\endgroup
\end{table*}

The reductions in Table~\ref{tab:resources} are direct circuit counts under nonlocal two-qubit connectivity and exclude routing and idle time; they leave the 50-CNOT source-syndrome-extraction stage unchanged.

The comparison has two diagnostic limitations. First, assigning noise to the seven final tomography readouts changes the circuit being diagnosed. Repeating the calculation for that 157-location diagnostic gives a cubic coefficient 423.58 and a 99.9\% upper bound $4.39\times10^{-7}$. Second, the advantage does not extend to minimum-weight error correction: the present leading coefficient is $183.81p^2$, compared with $25.5p^2$ in the published fit. The demonstrated advantage is therefore specific to the ideal-readout postselected diagnostic.

\section{Discussion and limitations}
\label{sec:discussion}
Specifying the stabilizer group does not specify the noisy preparation instrument.  Once checks share a flag circuit, the generating basis fixes which hooks appear, when they appear, and which later syndromes can still see them.  Bases that are equivalent in the ideal code thus leave open a physical fault-filtering degree of freedom.  In the present circuit that freedom suppresses the postselected logical error from quadratic to cubic order without adding a verification layer.  Exact order-two evaluation of 200 legal decompositions, including the reported basis, found 56 with $B_1=B_2=0$.  Because the reported basis was included a priori, this sample establishes nonuniqueness but estimates neither prevalence nor optimality.

The result applies to destructive logical tomography with error detection; it says nothing about the fidelity of a retained magic state. Postselection is a powerful fault-tolerance primitive~\cite{knill2005postselection,bombin2024fault}, but certifying the seven-qubit state while keeping it available would call for a nondestructive output check and a new fault analysis.  The present comparison also omits memory faults, routing, leakage, coherent error, crosstalk, and correlated noise.  Within those boundaries, stabilizer basis choice provides a circuit-level degree of freedom that can be explored in other flagged preparation and code-conversion protocols.

\section{Physical-platform feasibility}
\label{sec:platforms}
A 19-qubit footprint is modest for all three leading hardware modalities; the harder question is whether the 82 CNOTs, four reused ancillas, 23 mid-circuit measurements, and degree-seven verifier can be scheduled without erasing the resource advantage.  The component benchmarks quoted below come from different protocols and noise channels, so they speak to hardware compatibility and not to the postselected infidelity in Eq.~\eqref{eq:finite-certificate}.

\emph{Superconducting circuits.}  Local $T/T^\dagger$ rotations can be implemented as frame-tracked $Z$ rotations, and fast measurement and reset are mature.  The Willow system reports simultaneous mean errors of $0.035\%$ for one-qubit gates, $0.33\%$ for CZ gates, and $0.77\%$ for repeated measurement, together with a $1.1\,\mu\mathrm{s}$ surface-code cycle and real-time decoding~\cite{acharya2025below}.  The latest scalable-device calibration result reports a best CZ fidelity of $99.92(1)\%$ on an 84-qubit processor, but a $99.25\%$ median over 72 couplers~\cite{zhang2026cz}, so the best-case value cannot be assumed for every interaction in this circuit.  The main obstacle is topology rather than qubit count.  On a planar nearest-neighbor array, the seven verifier couplings and the grouped-check supports require SWAP routing, adding two-qubit gates, idle exposure, and leakage not present in the 82-CNOT model. A purpose-built 19-qubit patch with bus-mediated couplers, or a routing-aware redesign of the check order followed by a new fault enumeration, would be needed to preserve the quoted advantage.  Wall-clock time is shortest on this platform, but processor-wide entangling and readout errors should not yet be equated with the uniform $p=10^{-3}$ benchmark.

\emph{Neutral atoms.}  Reconfigurable tweezer arrays match the nonlocal check graph most naturally: data atoms can remain in storage while the four ancillas are moved through entangling and readout zones, avoiding coherent SWAP networks.  A 448-atom processor has already combined the same $[[15,1,3]]$ and $[[7,1,3]]$ code ingredients, programmable phase rotations, mid-circuit reuse, and transversal operations; it reports 270-ns CZ gates with fidelities up to $99.6\%$, together with nondestructive readout with $0.46(4)\%$ bit-flip error and $0.24(2)\%$ loss~\cite{bluvstein2026faulttolerant}. Separately, a 6100-qubit tweezer array reports $12.6(1)\,\mathrm{s}$ coherence, $99.99374(8)\%$ imaging fidelity, and approximately $99.95\%$ coherent transport fidelity~\cite{manetsch2025tweezer}.  The qubit count, connectivity, and the odd/even pattern of 15 local $T/T^\dagger$ rotations are all within reach.  The limiting operations are the sequential flag readouts and rearrangements; current CZ and nondestructive-readout errors also exceed the $10^{-3}$ model point, while atom loss supplies erasure information absent from our depolarizing analysis.  Routing overhead is smallest in this geometry, though a loss-aware circuit simulation is still needed before an output fidelity can be quoted.

\emph{Trapped ions.}  The 98-qubit Helios QCCD processor provides effective all-to-all connectivity, dynamic control, and native mid-circuit measurement and reset.  Averaged over its operation zones, reported infidelities are $2.5(1)\times10^{-5}$ for one-qubit gates, $7.9(2)\times10^{-4}$ for two-qubit gates, and $3.3(5)\times10^{-4}$ for state preparation and measurement; a maximally entangling gate takes about $70\,\mu\mathrm{s}$ and measurement-reset crosstalk on spectators is $1.3(1)\times10^{-5}$~\cite{ransford2026helios}. These numbers make ions the closest present match to the component-error scale used here, and ion transport eliminates logical SWAP gates for the high-degree checks.  The cost is time: the 82 entangling gates must be batched through a finite number of operation zones, producing transport, cooling, memory, and leakage exposure.  The same device also reports an effective mid-circuit measurement-reset error of $2.4(5)\times10^{-3}$ in random circuits.  Even this favorable mapping still needs a schedule-specific noise model and a repeat of the accepted-fault analysis. None of these component benchmarks establishes a live output fidelity, which lies outside the present claim in any case.

\section{Conclusion}
\label{sec:conclusion}

We have shown that the choice of stabilizer-generator basis can serve as an active design variable in fault-tolerant state preparation, rather than merely providing an algebraically equivalent description of the same code space. For the qRM-to-Steane conversion considered here, co-designing the generator basis with grouped flag measurements reshapes the syndrome information so that correlated faults generated across the non-Clifford layer are efficiently exposed to postselection. Under the ideal-readout destructive error-detection diagnostic, all accepted logical-error contributions from one- and two-fault events are eliminated, while the complete protocol requires only 19 qubits and 82 CNOT gates. The resulting postselected infidelity scales as \(210.2p^3+O(p^4)\), and exact low-order enumeration combined with finite-\(p\) sampling gives a 99.9\% upper bound of \(2.2\times10^{-7}\) at \(p=10^{-3}\).

The observed suppression is not an artifact of assuming perfect final tomography. When noise is also assigned to the seven final readout locations, the leading behavior remains cubic, with coefficient \(423.58\), and the corresponding 99.9\% upper bound at \(p=10^{-3}\) becomes \(4.39\times10^{-7}\). Thus, the elimination of accepted second-order faults survives the inclusion of noisy final measurements. The advantage is, however, diagnostic dependent. Under minimum-weight correction, the same circuit exhibits a leading contribution of \(183.81p^2\), compared with \(25.5p^2\) in the reference construction. The generator choice therefore improves the ability to detect and reject dangerous faults, but does not generically improve their correctability.

This distinction points to a broader design principle: stabilizer-generator optimization should be conditioned on the fault-tolerance task being performed. Generator bases that are advantageous for postselected preparation need not be optimal for active correction, decoding, or preservation of a live logical output. More generally, equivalent stabilizer descriptions provide a previously underused degree of freedom that can be optimized jointly with flag placement, measurement grouping, and code conversion. Extending this approach to nondestructive verification, other code-switching pairs, and architecture-specific noise models will determine whether generator-aware fault filtering can become a reusable tool for low-overhead preparation of non-Clifford resource states.

\section*{Data Availability Statement}
The numerical data supporting the findings of this study are available from the authors upon reasonable request.

\FloatBarrier
\appendix
\section{Subsystem conversion and bulk-\texorpdfstring{$X$}{X} gauge fixing}
\label{sec:codes}

Physical qubits are numbered from 1 to 15, and $P_S=\prod_{j\in S}P_j$.  We choose the seven-point hyperplane $H=\{1,\ldots,7\}$ as the retained Steane face and its complement $B=\{8,\ldots,15\}$ as the measured bulk. The logical representatives used throughout are
\begin{equation}
 \logical X_{\qRM}=X_{1\cdots7},\qquad
 \logical Z_{\qRM}=Z_{123},
\end{equation}
and, on $H$,
\begin{align}
 S_X^{\St}&=\{X_{1267},X_{2347},X_{4567}\},\nonumber\\
 S_Z^{\St}&=\{Z_{1267},Z_{2347},Z_{4567}\},\nonumber\\
 \logical X_{\St}&=X_{123},\qquad \logical Z_{\St}=Z_{123}.
\label{eq:steane}
\end{align}

To state the subsystem conversion precisely, define
\begin{align}
 h_1&=\{1,2,6,7\},&b_1&=\{8,9,13,14\},\nonumber\\
 h_2&=\{2,3,4,7\},&b_2&=\{9,10,11,14\},\nonumber\\
 h_3&=\{4,5,6,7\},&b_3&=\{11,12,13,14\},\nonumber\\
 &&b_4&=\{8,9,10,11,12,13,14,15\},
\label{eq:bottom-bulk-supports}
\end{align}
and $c_a=h_a\mathbin{\dot\cup}b_a$ for $a=1,2,3$, with $c_4=b_4$. The qRM code and the 15-qubit extended-Steane description share the rank-11 stabilizer center
\begin{equation}
 \begin{aligned}
 \mathcal S_0=\langle
 &X_{c_1},X_{c_2},X_{c_3},X_{c_4},\\
 &Z_{c_1},Z_{c_2},Z_{c_3},Z_{c_4},\\
 &Z_{h_1},Z_{h_2},Z_{h_3}\rangle .
 \end{aligned}
\label{eq:common-center}
\end{equation}
Thus $15-\operatorname{rank}\mathcal S_0=4$, so $\mathcal S_0$ defines a $[[15,4,3]]$ stabilizer parent~\cite{anderson2014conversion,quan2018conversion}. A canonical set of three gauge pairs is
\begin{align}
 (X_{g_1},Z_{g_1})&=(X_{1267},Z_{4,7,11,14}),\nonumber\\
 (X_{g_2},Z_{g_2})&=(X_{2347},Z_{4,5,11,12}),\nonumber\\
 (X_{g_3},Z_{g_3})&=(X_{4567},Z_{2,7,9,14}).
\label{eq:gauge-pairs}
\end{align}
They obey $X_{g_a}Z_{g_b}=(-1)^{\delta_{ab}}Z_{g_b}X_{g_a}$.  The two rank-14 gauge choices are
\begin{align}
 \mathcal S_{\qRM}&=\langle\mathcal S_0,Z_{g_1},Z_{g_2},Z_{g_3}\rangle,
 \nonumber\\
 \mathcal S_{\rm ERM}&=\langle\mathcal S_0,X_{g_1},X_{g_2},X_{g_3}\rangle.
\label{eq:two-gauges}
\end{align}
Accordingly, the precise subsystem notation is $[[15,1,3;3]]$: one of the four encoded degrees of freedom is protected and three are gauge.  Referring to a ``$[[15,4,3]]$ subsystem code'' is shorthand for the common $[[15,4,3]]$ stabilizer parent together with this protected/gauge split.

\subsection{Seven qubits collect all logical information after eight bulk measurements}
\label{sec:bulk}

In the direct single-register handoff used here, every qubit in $B$ is measured destructively in $X$.  This is not a measurement of eight independent gauge operators: only the three commuting $X_{g_a}$ are independent gauge observables.  Measuring all eight complementary physical qubits instead performs the gauge fix, disentangles and removes the bulk, and supplies an additional parity detector in one simultaneous readout without ancillas or CNOTs.

The logical-information statement follows directly in the $X$ gauge.  An equivalent generating set consists of the six Steane stabilizers on $H$ and the eight operators $\{X_{b_i},Z_{b_i}:i=1,\ldots,4\}$ on $B$.  The latter have full rank on eight qubits and therefore fix a unique bulk state, so
\begin{equation}
 \mathcal C_{\rm ERM}=\mathcal C_{\St,H}\otimes\ket{\Phi}_B.
\label{eq:erm-factorization}
\end{equation}
The unknown logical amplitudes occur only in the Steane factor.  Equivalently, $\logical Z_{\qRM}=\logical Z_{\St}=Z_{123}$ is supported entirely on $H$, while $\logical X_{\qRM}=X_{1\cdots7}=\logical X_{\St}X_{4567}$ and $X_{4567}$ is a Steane stabilizer.  Hence measuring or discarding $B$ cannot reveal or erase the protected logical information.

Write the eight recorded eigenvalues as $X_j=(-1)^{u_j}$ and define
\begin{align}
 s_1&=u_8\oplus u_9\oplus u_{13}\oplus u_{14},\nonumber\\
 s_2&=u_9\oplus u_{10}\oplus u_{11}\oplus u_{14},\nonumber\\
 s_3&=u_{11}\oplus u_{12}\oplus u_{13}\oplus u_{14}.
\label{eq:gauge-signs}
\end{align}
The retained state has $X_{h_a}$ eigenvalue $(-1)^{s_a}$.  A compact record-dependent frame is
\begin{equation}
 F(\bm u)=Z_{47}^{s_1}Z_{45}^{s_2}Z_{27}^{s_3},
\label{eq:compact-frame}
\end{equation}
which restores the three Steane $X$ checks and the logical-$X$ sign while leaving the logical-$Z$ operator unchanged.  If $K_{\bm u}$ projects the bulk onto the recorded $X$ eigenstates and deletes it, and $V_{\qRM},V_{\St}$ are encoding isometries for the protected qubit, then, up to a global phase,
\begin{equation}
 K_{\bm u}V_{\qRM}=
 \begin{cases}
  2^{-7/2}F(\bm u)V_{\St},&\displaystyle
       \bigoplus_{j=8}^{15}u_j=0,\\[2pt]
  0,&\displaystyle\bigoplus_{j=8}^{15}u_j=1.
 \end{cases}
\label{eq:gauge-kraus-map}
\end{equation}
This identity proves the logical channel record by record.  Ideally, all 128 even-parity records occur with probability $1/128$, independently of the input logical state.  The eight physical outcomes therefore decompose into one fixed-parity detector, three independent gauge/frame parities in Eq.~\eqref{eq:gauge-signs}, and four labels that distinguish equivalent measurement branches.  The three gauge signs are corrected by Eq.~\eqref{eq:compact-frame}; they are not postselection conditions.

Finally, let $A_{\rm src}$ indicate that the logical-$X$ verifier, all ten source syndromes, and all four source flags are trivial.  Let $g(\bm u)=\bigoplus_{j=8}^{15}u_j$ and let $\bm t\in\GF^3$ be the Steane syndrome.  The acceptance event used for the quoted diagnostic is
\begin{align}
 A_{\rm ED}&=A_{\rm src}\,\mathbf1[g(\bm u)=0]\,
                    \mathbf1[\bm t=\bm0],\nonumber\\
 P_{\rm ED}(p)&=\mathbb E_p[A_{\rm ED}],\qquad
 P_{\rm rej}(p)=1-P_{\rm ED}(p).
\label{eq:acceptance-indicator}
\end{align}
Thus the reported $P_{\rm ED}(10^{-3})\simeq86.90\%$ and the corresponding $13.10\%$ rejection are complementary probabilities for the full joint event, not for the bulk parity alone.  The source, bulk, and conditions are correlated, so the odd-bulk contribution cannot be inferred by subtraction.  In particular, requiring $s_1=s_2=s_3=0$ would be an incorrect extra postselection and would reduce the ideal acceptance to $1/8$.

\section{Grouped flag \texorpdfstring{$Z$}{Z}-check circuits}
\label{sec:gadgets}

The selected ten plaquettes are measured as two triples followed by two pairs,
\begin{equation}
 (p_{14},p_{16},p_{17})\longrightarrow(p_1,p_2,p_3)
 \longrightarrow(p_8,p_{12})\longrightarrow(p_{10},p_{11}).
\label{eq:block-order}
\end{equation}
They are distinct, have rank ten over $\GF$, and therefore span the complete qRM $Z$-stabilizer space.  The grouping changes the physical extraction instrument but not the ideal encoded state. Together with the logical-$X$ verifier, their ten syndrome bits and four flag bits form the 15 source-acceptance conditions.

For a pair, write $A=\{a,b,u_1,u_2\}$ and $B=\{a,b,v_1,v_2\}$. Syndrome ancillas $s_1,s_2$ start in $\ket0$ and a shared flag $f$ starts in $\ket+$. With $C_{r,s}=\mathrm{CNOT}_{r\rightarrow s}$, the ordered CNOT list is
\begin{align}
 U_2={}&C_{f,s_2}C_{f,s_1}C_{v_2,s_2}C_{v_1,s_2}
 C_{u_2,s_1}C_{u_1,s_1}\nonumber\\
 &\times C_{s_1,s_2}C_{b,s_1}C_{f,s_1}C_{a,s_1},
\label{eq:pair-sequence}
\end{align}
where the rightmost gate acts first.  Backward propagation gives
\begin{equation}
 U_2^\dagger(Z_{s_1},Z_{s_2},X_f)U_2
 =(Z_AZ_{s_1},Z_BZ_{s_1}Z_{s_2},X_f).
\label{eq:pair-observables}
\end{equation}
The final $Z$ measurements of $s_1,s_2$ therefore return the two requested check signs separately, and the ideal $X_f$ result is $+1$.

For a triple $A,B,C$, let $c$ lie in all three supports; let $d,e,g$ lie only in $A\cap B$, $A\cap C$, and $B\cap C$; and let $u_A,u_B,u_C$ be unique to one support, so $A=\{c,d,e,u_A\}$, $B=\{c,d,g,u_B\}$, and $C=\{c,e,g,u_C\}$.  The three syndrome ancillas and one shared flag use
\begin{align}
 U_3={}&C_{f,s_C}C_{f,s_B}C_{f,s_A}C_{g,s_C}C_{u_C,s_C}
 C_{e,s_C}C_{g,s_B}C_{u_B,s_B}\nonumber\\
 &\times C_{u_A,s_A}C_{e,s_A}C_{s_A,s_B}C_{d,s_A}
 C_{s_A,s_C}C_{f,s_A}C_{c,s_A},
\label{eq:triple-sequence}
\end{align}
again with the rightmost gate first.  Its ideal observables are
\begin{align}
 U_3^\dagger Z_{s_A}U_3&=Z_AZ_{s_A},\nonumber\\
 U_3^\dagger Z_{s_B}U_3&=Z_BZ_{s_A}Z_{s_B},\nonumber\\
 U_3^\dagger Z_{s_C}U_3&=Z_CZ_{s_A}Z_{s_C},\nonumber\\
 U_3^\dagger X_fU_3&=X_f.
\label{eq:triple-observables}
\end{align}
The actual checks and the geometry parameters that instantiate these two generic circuits are listed in Table~\ref{tab:grouped-checks}.

\begin{table}[t]
\caption{Ordered grouped checks and their circuit geometry.  Triple entries
are $(c,d,e,g;u_A,u_B,u_C)$; pair entries are
$(a,b;u_1,u_2;v_1,v_2)$.}
\label{tab:grouped-checks}
\centering
\begin{ruledtabular}
\begin{tabular}{cll}
group & ordered checks & geometry\\
\hline
$T_A$ & $(p_{14},p_{16},p_{17})$ & $(9,8,10,14;15,13,11)$\\
$T_B$ & $(p_1,p_2,p_3)$ & $(7,2,6,4;1,3,5)$\\
$P_A$ & $(p_8,p_{12})$ & $(4,11;5,12;7,14)$\\
$P_B$ & $(p_{10},p_{11})$ & $(7,14;6,13;2,9)$
\end{tabular}
\end{ruledtabular}
\end{table}

The simultaneous shared-flag topologies originate in Ref.~\cite{chao2018quantum} and were adapted to qRM plaquettes in Ref.~\cite{daguerre2025code}; the particular generator basis in Table~\ref{tab:grouped-checks} is selected here.  The reason for the flag follows from two CNOT conjugation rules,
\begin{equation}
 C_{d,s}Z_sC_{d,s}=Z_dZ_s,
 \qquad C_{f,s}Z_sC_{f,s}=Z_fZ_s.
\label{eq:flag-propagation}
\end{equation}
A $Z_s$ component on a syndrome target can copy through later data-to-syndrome CNOTs and create a correlated data-$Z$ hook.  The same component crosses a flag-to-syndrome CNOT and becomes $Z_f$, which flips the final $X_f$ result.  The opening and closing flag couplings in Eqs.~\eqref{eq:pair-sequence} and \eqref{eq:triple-sequence} place each single-ancilla fault interval capable of producing an uncorrectable hook in a nontrivial flag sector, as required by the standard flag criterion~\cite{chamberland2018flag,tansuwannont2020flag}.

Grouping has two roles.  First, one flag protects two or three syndrome rails, so ten checks require four grouped gadgets rather than ten separate flag rounds; the three syndrome ancillas and one flag can be reset and reused. Second, the generator basis and group order determine the detector signatures of the hooks.  Changing the basis leaves the stabilizer space unchanged but changes which faults are exposed by later groups, the bulk parity, and the Steane checks.  The local flag criterion is only a one-fault statement; it does not by itself prove the reported two-fault cancellation. That stronger property was checked by complete propagation through the transversal layer, gauge frame, and detector:
\begin{equation}
 |F|\le2,\qquad D(F)=0\quad\Longrightarrow\quad\ell(F)=0,
\label{eq:zero-b2-test}
\end{equation}
which gives $B_1=B_2=0$ for the basis and order in Table~\ref{tab:grouped-checks}.

For scope, the fixed triple--triple--pair--pair template admits 36 legal triples and 60 legal pairs among the 18 qRM plaquettes.  Requiring ten distinct rank-ten checks gives 71,232 ordered candidates.  The reported basis was included a priori in a 200-candidate order-two evaluation, in which 56 bases obeyed Eq.~\eqref{eq:zero-b2-test}.

\section{Low-order evaluation and finite-\texorpdfstring{$p$}{p} certificate}
\label{sec:dkm-exact-finite}

The matched uniform model assigns probability $p$ to each preparation or live measurement flip, one-qubit depolarizing noise after each physical $T/T^\dagger$, and two-qubit depolarizing noise after each CNOT; idle errors are absent.  The seven final Steane measurements are ideal destructive tomography operations, not noisy locations.  The diagnostic therefore has
\begin{equation}
 \begin{split}
 N_{\rm loc}={}&30\ \text{preparations}+82\ \text{CNOTs}\\
 &+15\ T/T^\dagger+23\ \text{measurements}=150.
 \end{split}
\label{eq:dkm-nloc}
\end{equation}

Pauli faults crossing $T/T^\dagger$ can acquire local $S$ or $S^\dagger$ factors, so the calculation retains the complete physical Clifford frame. The target state is evaluated with the four-branch stabilizer decomposition
\begin{equation}
 \begin{split}
 \proj{T}={}&\frac{1+\sqrt2}{4}(\proj{+}+\proj{Y+})\\
 &+\frac{1-\sqrt2}{4}(\proj{-}+\proj{Y-}).
 \end{split}
\label{eq:t-decomposition}
\end{equation}
For branch $r$ and basis $b\in\{X,Y\}$, let $W_{r,b}$ be the accepted weight and $V_{r,b}$ the accepted logical numerator.  The signed branch sums are formed before conditional normalization,
\begin{align}
 W_b&=\sum_r\alpha_rW_{r,b},&V_b&=\sum_r\alpha_rV_{r,b},\nonumber\\
 P_{\rm ED}&=\frac{W_X+W_Y}{2},&
 Q_{\rm ED}&=\frac{W_X+W_Y}{4}
 -\frac{V_X+V_Y}{2\sqrt2},\nonumber\\
 I_{\rm ED}&=Q_{\rm ED}/P_{\rm ED}.
\label{eq:dkm-contract}
\end{align}
Normalizing the four branches separately would be incorrect because the last two coefficients in Eq.~\eqref{eq:t-decomposition} are negative.

Exact subset enumeration with stabilizer-tableau propagation ~\cite{aaronson2004simulation,gidney2021stim} gives
\begin{equation}
 \begin{split}
 I_{\rm ED}(p)&=\frac{2837363}{13500}p^3+O(p^4)\\
 &=210.175p^3+O(p^4),\qquad B_1=B_2=0.
 \end{split}
\label{eq:dkm-exact-series}
\end{equation}
Orders zero through three are evaluated exactly.  At $p=10^{-3}$, orders four, five, and six use $10^6$ uniformly weighted fixed-order samples each. For a sample loss $\ell_\omega$, the one-sided construction uses
\begin{equation}
 \ell_\omega\le\mathbf1[\ell_\omega>0]\le1;
\label{eq:event-bound}
\end{equation}
negative signed losses can only lower the desired upper bound.  A one-sided Clopper--Pearson bound~\cite{clopper1934binomial} is applied to the positive event rate in each stratum, with a Bonferroni allocation of the total $10^{-3}$ error probability.  Table~\ref{tab:finite-strata} gives the only sampled inputs required for the certificate.

\begin{center}
\refstepcounter{table}\label{tab:finite-strata}
\begin{minipage}{\columnwidth}
\small\textbf{TABLE \thetable.} Fixed-fault strata for the finite-$p$ certificate at $p=10^{-3}$.  The last column is the upper contribution to unnormalized loss.
\medskip
\centering
\begin{ruledtabular}
\begin{tabular}{crrr}
order & samples & positive events & loss upper\\
\hline
4 & $10^6$ & 292 & $6.21\times10^{-9}$\\
5 & $10^6$ & 199 & $1.29\times10^{-10}$\\
6 & $10^6$ & 121 & $2.02\times10^{-12}$
\end{tabular}
\end{ruledtabular}
\end{minipage}
\end{center}

The exact cubic contribution to unnormalized loss is $1.8143\times10^{-7}$, and all orders seven and higher are assigned unit loss, giving a tail below $2.60\times10^{-10}$.  Combining the exact term, the three simultaneous bounds, and the adversarial tail gives $Q_{\rm ED}^{\rm up}=1.8803\times10^{-7}$.  The zero-fault event supplies the denominator floor $P_{\rm ED}\ge(1-p)^{150}$, hence
\begin{equation}
 I_{\rm ED}(10^{-3})\le2.19\times10^{-7}
 \qquad(99.9\%\ \text{joint confidence}).
\label{eq:dkm-finite-upper}
\end{equation}
The accepted-weight expansion through third order, together with the entire probability mass of four or more faults, gives
\begin{equation}
 0.8689553\le P_{\rm ED}(10^{-3})\le0.8689734.
\label{eq:dkm-acceptance-interval}
\end{equation}
This is the interval summarized as $86.90\%$ in the main text and in Eq.~\eqref{eq:acceptance-indicator}.  It uses ideal tomography; the separate diagnostic in which the seven final readouts are noisy is not part of the matched headline result.

\FloatBarrier
\nocite{apsrev42Control}
\bibliographystyle{apsrev4-2}
\renewcommand{\bibfont}{\footnotesize}
\setlength{\bibsep}{0pt}
\bibliography{reference}

\end{document}